\documentclass[acmlarge]{acmart}

    \usepackage{color, colortbl}
\usepackage{enumitem}
\usepackage[toc,page]{appendix}
\usepackage{graphicx}
\usepackage{caption}
\usepackage{subcaption}
\usepackage{tabularx}
\usepackage{wrapfig}
\usepackage{pdfpages}
\usepackage{tikz}

\begin{document}

% \title{Recommend Greener Items: Towards Carbon Footprint-Aware Recommender Systems}
\title{Towards Carbon Footprint-Aware Recommender Systems for Greener Item Recommendation}
%\subtitle{A Dataset, Benchmark Analysis and Reranking Approach}
%\subtitle{Dataset, Benchmarking and Reranking}

%\subtitle{Empowering Recommender Systems for Carbon Footprint-Aware Recommendations}

%Carbon Footprint-Aware Recommender Systems: A Benchmark Analysis and Reranking Approach

\author{Raoul Kalisvaart}
\email{contact@raoulkalisvaart.com}
%\email{webmaster@marysville-ohio.com}
\affiliation{%
  \institution{Delft University of Technology}
 % \streetaddress{P.O. Box 1212}
  \city{Delft}
  \country{The Netherlands}
}

\author{Masoud Mansoury}
\email{m.mansoury@tudelft.nl}
%\email{webmaster@marysville-ohio.com}
\affiliation{%
  \institution{Delft University of Technology}
 % \streetaddress{P.O. Box 1212}
  \city{Delft}
  \country{The Netherlands}
}

\author{Alan Hanjalic}
\email{a.hanjalic@tudelft.nl}
%\email{webmaster@marysville-ohio.com}
\affiliation{%
  \institution{Delft University of Technology}
 % \streetaddress{P.O. Box 1212}
  \city{Delft}
  \country{The Netherlands}
}

\author{Elvin Isufi}
\email{e.isufi-1@tudelft.nl}
%\email{webmaster@marysville-ohio.com}
\affiliation{%
  \institution{Delft University of Technology}
 % \streetaddress{P.O. Box 1212}
  \city{Delft}
  \country{The Netherlands}
}

\input{mysymbol.sty}

\renewcommand{\shortauthors}{Kalisvaart, Mansoury, Hanjalic, Isufi}

\begin{abstract}
%The growing awareness of the environmental impact of e-commerce demands more sustainable recommendations to address this societal aspect and satisfy purpose-driven users. However, the role of recommender systems (RecSys) in promoting sustainable shopping remains unexplored mainly because we lack a dataset containing carbon footprint emissions for the items. In this paper, we address this bottleneck and provide a first study on the environmental role of RecSys algorithms. We first present a dataset that includes carbon footprint emissions for its items. Then, we benchmark conventional RecSys algorithms in terms of accuracy and sustainability, establishing an accuracy-greenness tradeoff. We find that RecSys algorithms optimized for accuracy overlook greenness, and that longer recommendation lists are greener but less accurate. Finally, we propose a reranking algorithm to improve this tradeoff. This reranking approach is modular and applies to all RecSys algorithms without altering their underlying mechanisms or even without needing a retrain of them. Our results demonstrate that sacrificing a small degree of accuracy can lead to significant improvements in the recommendation greenness across all algorithms and list lengths. We anticipate that this work will serve as the starting point to study RecSys in promoting sustainable e-commerce and researching all facets this new direction poses. %Additionally, the methods proposed in this study could prove to be useful in addressing climate challenges beyond e-commerce, such as sustainable transportation or commuting alternatives.
%
The commodity and widespread use of online shopping are having an unprecedented impact on climate, with emission figures from key actors that are easily comparable to those of a large-scale metropolis. Despite online shopping being fueled by recommender systems (RecSys) algorithms, the role and potential of the latter in promoting more sustainable choices is little studied. One of the main reasons for this could be attributed to the lack of a dataset containing carbon footprint emissions for the items. While building such a dataset is a rather challenging task, its presence is pivotal for opening the doors to novel perspectives, evaluations, and methods for RecSys research. In this paper, we target this bottleneck and study the environmental role of RecSys algorithms. First, we mine a dataset that includes carbon footprint emissions for its items. Then, we benchmark conventional RecSys algorithms in terms of accuracy and sustainability as two faces of the same coin. We find that RecSys algorithms optimized for accuracy overlook greenness and that longer recommendation lists are greener but less accurate. Then, we show that a simple reranking approach that accounts for the item's carbon footprint can establish a better trade-off between accuracy and greenness. This reranking approach is modular, ready to use, and can be applied to any RecSys algorithm without the need to alter the underlying mechanisms or retrain models. Our results show that a small sacrifice of accuracy can lead to significant improvements of recommendation greenness across all algorithms and list lengths. Arguably, this accuracy-greenness trade-off could even be seen as an enhancement of user satisfaction, particularly for purpose-driven users who prioritize the environmental impact of their choices. We anticipate this work will serve as the starting point for studying RecSys for more sustainable recommendations.
\end{abstract}

%\begin{CCSXML}
%<ccs2012>
%  <concept_id>10010520.10010553.10010562</concept_id>
%  <concept_desc>Computer systems organization~Embedded systems</concept_desc>
%  <concept_significance>500</concept_significance>
%  <concept_id>10010520.10010575.10010755</concept_id>
%  <concept_desc>Computer systems organization~Redundancy</concept_desc>
%  <concept_significance>300</concept_significance>
%  <concept_id>10010520.10010553.10010554</concept_id>
%  <concept_desc>Computer systems organization~Robotics</concept_desc>
%  <concept_significance>100</concept_significance>
%  <concept_id>10003033.10003083.10003095</concept_id>
%  <concept_desc>Networks~Network reliability</concept_desc>
%  <concept_significance>100</concept_significance>
%</ccs2012>
%\end{CCSXML}

%\ccsdesc[500]{Computer systems organization~Embedded systems}



\maketitle

%%%%%%
%	Introduction
\section{Introduction}\label{sec:introduction}

The carbon emissions resulting from online shopping keep increasing, despite the promises of the main actors to adopt long-term carbon-neutral practices~\cite{AmazonStatist, AliBabaStatistSource, WEFCarbon}. For instance, the Chinese e-commerce was responsible for $\sim48$ million metric tons (MMT) of carbon dioxide equivalent (CO$_2$-eq) in 2019, and this figure is expected to rise to 105 MMT in 2025 \cite{CarbonStop}. \footnote{CO$_2$-eq is a universal measure of the amount of greenhouse gases produced that have the same global warming potential as CO$_2$~\cite{eucarbon}.} Amazon also reported more than 71 MMT of CO$_2$-eq emissions in 2021, and by simple linear forecasting, it will likely follow the prediction of the Chinese e-commerce. To provide some perspective, The Netherlands -- the 15th largest economy in the world in terms of gross domestic product per capita \cite{Forbes} -- produced 178.2 MMT of CO$_2$-eq in 2021, and its emission data from the last 16 years show a decreasing trend.

% Given that e-commerce nowadays is driven by recommender system (RecSys) algorithms \cite{schafer1999recommender, wei2007survey, aggarwal2016recommender}, the questions that are becoming more and more relevant are what role do RecSys play in the overall e-commerce carbon footprint landscape and to what extent they can contribute to reducing this footprint.
Given the central role of recommender system (RecSys) algorithms in e-commerce \cite{schafer1999recommender, wei2007survey, aggarwal2016recommender}, it is increasingly important to understand their role towards shaping the sector's carbon footprint—and to what extent they can help reduce it. Answering these questions is complex as it requires a deep analysis of the entire pipeline from item (product) production, via distribution and logistics, to delivery, and the role of RecSys in different segments of this pipeline; as we extensively review in Sec.~\ref{sec:relatedwork}. For example, RecSys could focus more closely on the carbon footprint of the production of items and take this as an extra criterion on what to recommend to users. However, they could also consider minimizing the delivery effort (distance, time, fuel consumption) in bringing the items to the users, e.g., by recommending similar items but from a closer distribution center. Furthermore, RecSys could focus on providers with a proven sustainability track record and match the products they offer to those the users prefer. Finally, incorporating sustainability into recommendations could diversify options and increase serendipity, broadening users’ preferences toward more sustainable choices and encouraging a positive shift in mindset. All of this calls for a complex, holistic, and multi-criterion analysis in which sustainability is accounted for, while keeping the users loyal to a recommendation platform. In addition, this calls for data that, next to the information on user and item properties and user preferences, contain information on the sustainability aspects. The absence of this factor is acknowledged as the primary bottleneck limiting the study of RecSys algorithms in addressing climate change~\cite{rolnick2022tackling}.

When associating sustainability with recommender systems, the conventional approach is to focus on the carbon emissions generated by the recommendation algorithms during the training and operation phases~\cite{spillo2023towards}.
%Research on sustainable recommender systems has typically focused on the carbon emissions generated by the recommendation algorithms during the training and operation phases~\cite{spillo2023towards}. 
While these studies address sustainability from the perspective of the computational costs associated with model training and execution, our paper introduces a different dimension: \textit{recommending greener items}. In this context, a recommender system is considered sustainable if it prioritizes items that require less carbon emission from production to delivery. We illustrate this approach with two examples from distinct domains:

\begin{itemize}
    \item \textbf{Product Recommendation in E-Commerce:} Consider a scenario where a recommender system identifies two products that match a user's preferences. Item A is located overseas, while Item B is produced and located in the proximity. Although both items are equally relevant to the user's preferences, delivering Item A would generate higher carbon emissions due to the greater shipping distance. With knowledge of the carbon footprint associated with each delivery, a sustainable recommendation model would favor recommending Item B over Item A, thereby minimizing carbon emissions.
    \item \textbf{Recipe Recommendation in the Food Market:} In another example, a recommendation model identifies two recipes that align with a user's preferences. Recipe A contains ingredients that have a high carbon footprint due to their production processes, while Recipe B uses ingredients with a much lower carbon footprint. Given that both recipes are equally relevant to the user, a sustainable recommendation model would prioritize Recipe B, which requires less carbon emission. This approach not only promotes greener food choices but also potentially encourages more sustainable food production and consumption by increasing the demand for low-carbon foods.

\end{itemize}

Research on this new aspect of sustainable recommender systems requires datasets that include specific information, such as the delivery distance of items to users (e.g., product recommendation in e-commerce) or the carbon emissions associated with the production of items (e.g., recipe recommendation in the food market). However, such datasets are currently unavailable. Since it is impossible to cover all sustainability aspects in one framework, some of which require an extensive interdisciplinary effort, we focus on the specific goal of recommending \emph{greener} items, where their carbon footprint plays a critical role when assessing their suitability for recommendation. More specifically, given the  availability of online APIs that provide carbon footprint data for food ingredients, we concentrate on the second example—recipe recommendations in the food market. Consequently, we look at the context of recommending cooking recipes to users, in which, next to the match to a user’s taste, the carbon footprint of the used ingredients plays a role. While this domain-specific focus is chosen to retrieve the necessary carbon footprint data for items, our principal goal does not revolve around food recommendations or their nutritional aspects, which have been extensively studied elsewhere~\cite{ge2015using,wagner2014spatial}. Instead, the aim is to leverage such datasets to shed more light on the possibilities of empowering RecSys to recommend green choices. We break down this objective by answering the following research questions:
\begin{enumerate}[label=(RQ\arabic*),start=1]
    \item How does our newly developed dataset, which includes recipes and their associated greenness scores, compare with typical benchmark datasets used for recommender systems research in terms of sparsity, engagement, and data distributions?
    \item Do conventional recommendation algorithms recommend greener items for top$-k$? 
    \item How can we make conventional recommendation algorithms recommend greener items without altering their inner-working mechanisms?
\end{enumerate}
In view of this, we make the following threefold contribution:

\begin{enumerate}[label=\emph{Answer to RQ\arabic*:},start=1]
\item We construct the RecipeEmission dataset, the first RecSys dataset that includes carbon footprint emissions for the items as features. RecipeEmission contains interactions between users and cooking recipes, consisting of 32,093 users, 5,605 items, having a sparsity of 99.86\%. Each item (recipe) contains its CO$_2$-eq footprint and greenness score. RecipeEmission exhibits a sparsity and long-tail properties similar to conventional RecSys datasets (Fig.~\ref{fig:dataset_props} and Table~\ref{tab_stat}). The dataset can be accessed at \url{https://github.com/RaoulKalisvaart/green-recommender-systems}.
 \item We conduct a comprehensive benchmark analysis on nine conventional RecSys algorithms in terms of the trade-off between accuracy and greenness. In this regard, we propose a metric to measure the greenness of a list by following the normalized discounted cumulative gain rationale. Our analysis reveals that RecSys algorithms are not biased toward greener choices by default, requiring innovation towards developing RecSys for a better balance between recommending green items and satisfying user preferences. More specifically, all algorithms ignore greenness in their recommendations, and longer top-$k$ lists are greener but less accurate. We also find that model-based algorithms, such as SVD, provide a better trade-off than neighbor-based alternatives such as nearest neighbors (Fig.~\ref{fig:co_rank_rating} and Table~\ref{tab_ratPerf}).
 \item To empower RecSys towards recommending greener items, we resort to a simple yet effective reranking strategy. It considers a green-accuracy utility score to update the recommendation list and produce a better accuracy-greenness trade-off. This approach is model-agnostic and ready-to-use approach, familiar to the RecSys community, and can be easily tested in any existing and trained RecSys algorithm. We find that it is possible to sacrifice a small degree of accuracy to achieve a substantial improvement in greenness, as well as a better trade-off for shorter lists and factorization-based methods (Fig.~\ref{fig:tradeoff_list_lengths}).
\end{enumerate}

This paper is structured as follows. Sec.~\ref{sec:problem} details the problem studied in this paper, whereas Sec. \ref{sec:data} describes the properties and the methods to build it. Sec. \ref{sec:results} first benchmarks nine conventional RecSys algorithms in terms of accuracy and greenness, and then it provides the reranking-based method to update the recommendation list based on both greenness and relevance. Sec. \ref{sec:discussion} provides a broad discussion on the findings. %Methods are deferred to Sec. \ref{sec:methods} to make the paper accessible to a broader audience, where we detail the challenges of building the dataset and the experimental setup to assess the RecSys algorithms. 
Sec.~\ref{sec:relatedwork} reviews the related works, user studies, and surveys on the broader impact of carbon footprint in e-commerce and the role of RecSys algorithms in all this. The paper is concluded in Sec.~\ref{sec:conclusion}.

\section{Problem formulation}\label{sec:problem}

We consider the recommendation task of retrieving the top-$k$ items with the highest predicted relevance scores for each target user. Let $U$ denote the set of users and $\mathcal{I}$ the set of items. We define $\bbR$ as the user-item rating matrix, where each entry of this matrix represents the rating given by a user $u$ to an item $i$. The recommendation algorithm uses $\bbR$ as input and generates recommendation lists of size $k$ for each user. %To study the greenness of recommendation models (i.e., recommending greener items to the users), we associate a \textit{greenness} value to each item indicating how green the item is or in contrary how much carbon emission is involved in producing or delivering this item. We denote the greenness value of an item $i$ as $g_i$.
We associate to each item $i$ a \emph{greenness value}, $g_i$, that indicates how environmentally friendly item $i$ is, whether in terms of carbon emissions involved in its production, delivery, or other sustainability factors. The definition of item greenness varies across domains as discussed in the two examples on e-commerce and food market in the previous section.%

%. For example, in e-commerce, the delivery distance from seller to buyer can determine greenness, as shorter distances result in lower carbon emissions. %Given equally relevant items, those requiring less transportation would sore higher. 
%As another example, in a food recommendation domain, the greenness of a food item can be defined based on the carbon footprint of its ingredients. Recipes with lower-carbon ingredients are considered greener and should be favored when relevance scores are equal.
%}

%\textcolor{blue}{To assess the sustainability and greenness of recommendation models--specifically, their ability to prioritize greener items--we associate each item with a \textbf{greenness value}, which quantifies its environmental impact. This value, denoted as $g_i$ for an item $i$, indicates how environmentally friendly an item is, whether in terms of carbon emissions involved in its production, delivery, or other sustainability factors. However, this greenness value is not used as an input to the recommendation model itself. Instead, it serves two purposes: (1) assessing how well existing models prioritize greener items, and (2) integrating greenness into recommendation results via a post-processing approach for enhancing recommendation greenness.}

In this work, the item greenness value is not used as an input to the recommendation model itself. Instead, it serves two purposes: (1) assessing how well existing models prioritize greener items over alternatives, and (2) integrating greenness into recommendation results via a post-processing approach for enhancing recommendation greenness. Ideally, given two items, $A$ and $B$ with equal relevance scores but different greenness values (e.g. $g_A=0.7$ (higher greenness) and $g_B=0.3$ (lower greenness)), a sustainability-aware recommendation model should prioritize $A$ over $B$. Beyond evaluating the greenness of existing models, we propose a reranking approach to explicitly enhance recommendation sustainability by adjusting rankings to favor greener items while maintaining overall accuracy.

%\textcolor{blue}{The definition of item greenness varies across domains. For example, in e-commerce, the delivery distance from seller to buyer can determine greenness, as shorter distances result in lower carbon emissions. Given equally relevant items, those requiring less transportation should be prioritized. As another example, in a food recommendation domain, the greenness of a food item can be defined based on the carbon footprint of its ingredients. Recipes with lower-carbon ingredients are considered greener and should be favored when relevance scores are equal.}

%\textcolor{blue}{The primary goal of this research is to investigate how well existing recommendation models prioritize greener items over alternatives. Ideally, given two items, $A$ and $B$ with equal relevance scores but different greenness values as $g_A=0.7$ (higher greenness) and $g_B=0.3$ (lower greenness), a sustainability-aware recommendation model should prioritize $A$ over $B$. Beyond evaluating existing models, we propose a reranking approach to explicitly enhance recommendation sustainability by adjusting rankings to favor greener items while maintaining the overall accuracy.}

Clearly, the key challenge in conducting this research is to identify the greenness values $g_i$ for the items in a suitable RecSys dataset. To this end, we consider the food recommendation domain, and collect the greenness information using online APIs that provide carbon footprint data for ingredients constituting a recipe. We start with an existing, readily available dataset that lacks greenness values and enhance it with such information to support research in sustainable recommendation. Although our conclusions are dataset- and domain-biased, the proposed methodology, insights, and broader perspective is generalizable to other domains, provided that greenness data is available. %The recommendation models themselves do not require greenness information for training, but we incorporate it into the evaluation process to measure sustainability.

%\textcolor{blue}{A key challenge in conducting this research is the lack of existing recommendation datasets that include greenness information. To address this, we prepare and release a new dataset in the food recommendation domain, where greenness information is collected using online APIs that provide carbon footprint data for food ingredients. While the dataset does not originally contain greenness values, we augment it with this information to enable research in sustainable recommendation. Although our study focuses on food recommendations, the proposed methodology is generalizable to other domains, provided that greenness data is available. The recommendation models themselves do not require greenness information for training, but we incorporate it into the evaluation process to measure sustainability.}
%%%%%%

%	DATA
\section{R\lowercase{ecipe}E\lowercase{mission} Data}\label{sec:data}

\begin{wrapfigure}{r}{0.5\textwidth}
\vskip-.5cm
\begin{minipage}{80mm}
  % \centering
  \raggedleft
\begin{align*}
    % \begin{split}
         & \texttt{user id:} ~~~~56,680&\\
         & \texttt{rating:} ~~~~5&\\
         & \texttt{date:} ~~~~2006-11-11&\\
         & \texttt{review:} ~~~~\texttt{Great!}&\\
         & \texttt{recipe id:} ~~
            \begin{cases}
              \texttt{recipe name:} & \texttt{patato}\\
              \texttt{ingredient ids:} & [2851,\ldots]\\
              \texttt{CO$_2$ footprint:} & 4.46 \\
              \texttt{ingredient names:} & [\texttt{onion},\ldots]\\
              \texttt{ingredient quantities:} & [1,\ldots] \\
              \texttt{greenness:} & 3.37
            \end{cases}  &
    % \end{split}
\end{align*}
\captionof{figure}{An example of an entry in the RecipeEmission dataset. }
        \label{recipe_entry}
\end{minipage}
  \vskip-.5cm
\end{wrapfigure}

%PREAMBLE
In this section, we first analyze the developed RecipeEmission dataset and answer RQ1. Then, we describe the methods for building this dataset. %Our initial data come from the Food.com dataset in \cite{majumder2019generating}, which contains users interacting with recipe items via explicit ratings and textual reviews. Each recipe has its name and list of ingredients as a feature. As additional features, we mined the ingredient quantities, \emph{the CO$_2$ equivalent} (CO$_2$-eq) of the recipes and their greenness. Greenness is a quantization of the CO$_2$-eq into a finite scale of real-valued scores. An example of an entity is shown in Fig.~\ref{recipe_entry}. %In this section, we first present the steps followed to build the RecipeEmission dataset in Sec.~\ref{subsec:buildData}.

%SUBSECTION
\subsection{Dataset Analysis} 

\smallskip
\noindent\textbf{Statistics.}
RecipeEmission has $32.093$ users, $5.605$ recipes, and $247.521$ interactions, modelled as explicit ratings $\{0,\ldots,5\}$ leading to a sparsity of $99.86\%$. Each recipe has its name and list of ingredients as a feature. As additional features, we mined the ingredient quantities, \emph{the CO$_2$ equivalent} (CO$_2$-eq) of the recipes and their greenness. Greenness is a quantization of the CO$_2$-eq into a finite scale of real-valued scores. An example of an entity is shown in Fig.~\ref{recipe_entry}.
%

%%%%%%%%%%%%%%%%%%%%%%%%%%%%%%%%%%%%%%%%%%%%%%%%%%%%%%%%%%%%%%%%%%%%%%%%%%%%%%%%%%%%%
%%%%%%%%%%%%%%%%%%%%%%%%%%%%%%%%%%%%%%%%%%%%%%%%%%%%%%%%%%%%%%%%%%%%%%%%%%%%%%%%%%%%%
%   STATISTIC TABLE
\begin{figure*}[!tp]
\centering
\begin{subfigure}{.33\textwidth}
  \centering
  \includegraphics[width=1\linewidth]{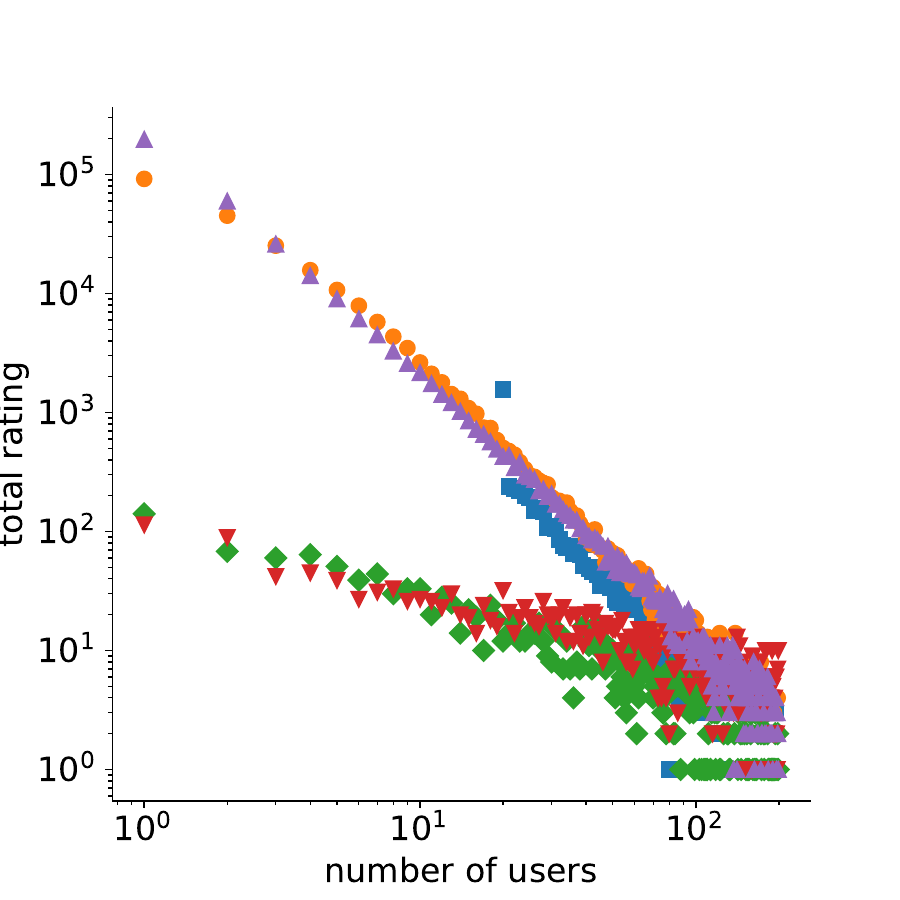}
  \caption{}
  \label{fig:sub_user_rank}
\end{subfigure}%
\begin{subfigure}{.33\textwidth}
  \centering
  \includegraphics[width=1\linewidth]{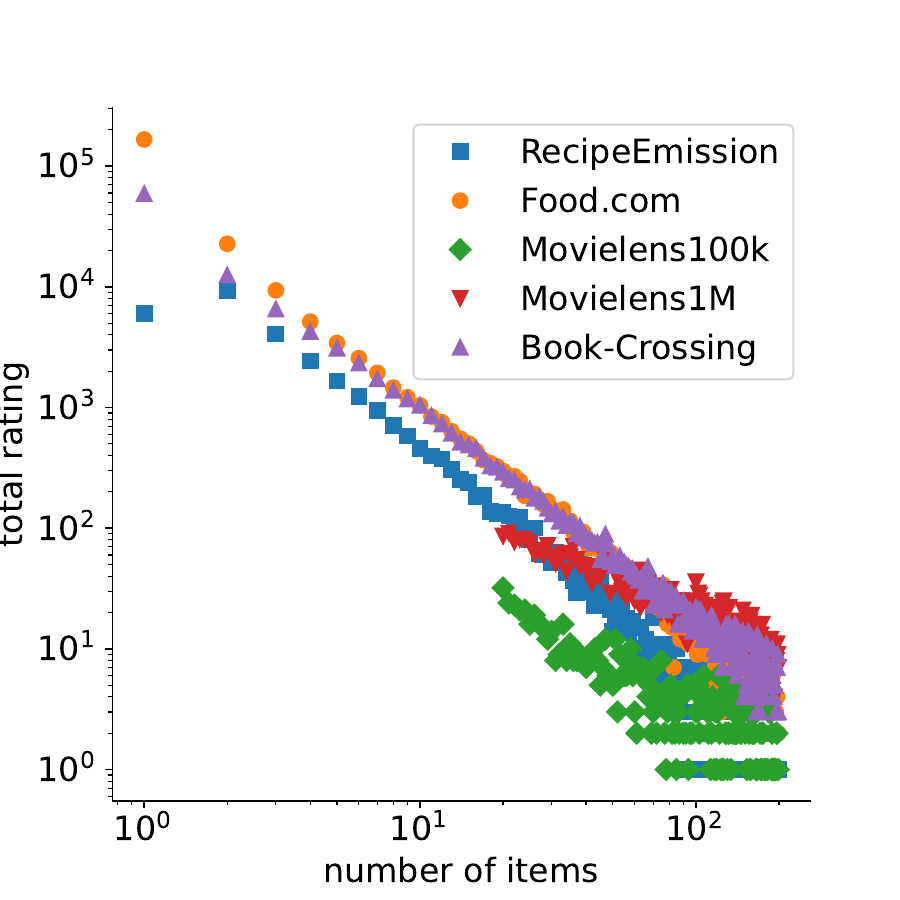}
  \caption{}
  \label{fig:sub_item_rank}
\end{subfigure}
\begin{subfigure}{.33\textwidth}
\centering   \includegraphics[width=1\linewidth]{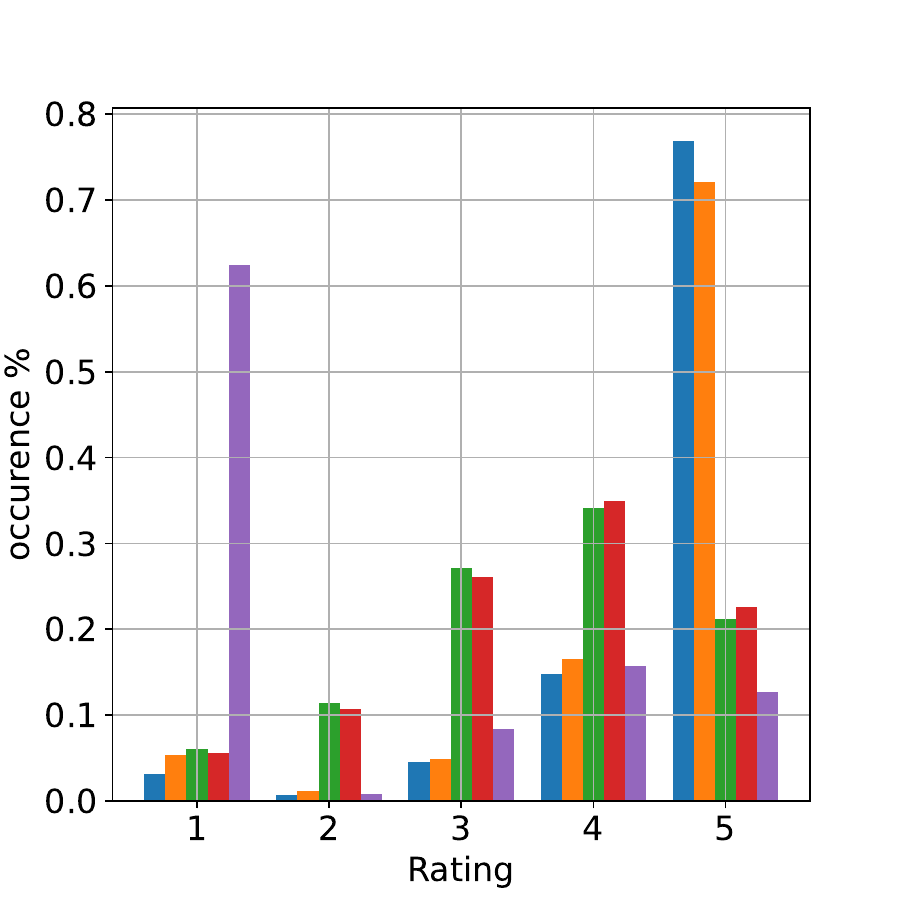}
\caption{}
\end{subfigure}
%
%\caption{Long-tail distributions of the RecipeEmission and other datasets. (a) User engagement distribution (total ratings of each user). The RecipeEmission dataset has a similar distribution as the starting Food.com dataset and baselines. There are no users with a large number of ratings because of the filtration steps over the minimum number of interactions per item. (b) Item popularity distribution (total ratings of each item). The distribution of the RecipeEmission dataset follows closely that of the original Food.com. MovieLens datasets filter on the minumum number of interactions per user which can be noted by the trunkated distributions in this panel.}
\caption{Distribution of the statistics for the RecipeEmission and conventional datasets. \textbf{(a)} Long-tail distributions of the user engagement (total ratings of each user). RecipeEmission has a similar distribution as the starting Food.com dataset and baselines. There are no users with a large number of ratings because of the filtration steps over the minimum number of interactions per item. \textbf{(b)} Long-tail distributions of the item popularity distribution (total ratings of each item). RecipeEmission closely follows the distribution of the original Food.com. MovieLens datasets are filtered on the minimum number of interactions per user which can be noted by the truncated distributions in this panel. \textbf{(c)} Rating distribution scaled to the set $\{1, \ldots, 5\}$. All datasets but Book-Crossing have a skew towards high rating values. RecipeEmission respects the distribution of the original Food.com dataset, which follows a similar pattern as the MovieLens datasets.}
\label{fig:dataset_props}
\end{figure*}

%%%%%%%%%%%%%%%%%%%%%%%%%%%%%%%%%%%%%%%%%%%%%%%%%%%%%%%%%%%%%%%%%%%%%%%%%%%%%%%%%%%%%
%%%%%%%%%%%%%%%%%%%%%%%%%%%%%%%%%%%%%%%%%%%%%%%%%%%%%%%%%%%%%%%%%%%%%%%%%%%%%%%%%%%%%

%
Figs.~\ref{fig:dataset_props}(a) and~\ref{fig:dataset_props}(b) show that RecipeEmission preserves key properties of conventional RecSys datasets such as the long-tail distribution of both user engagement --most user engage with few items-- and item popularity --a few items collect most ratings.

%%%%%%%%%%%%%%%%%%%%%%%%%%%%%%%%%%%%%%%%%%%%%%%%%%%%%%%%%%%%%%%%%%%%%%%%%%%%%%%%%%%%%
%%%%%%%%%%%%%%%%%%%%%%%%%%%%%%%%%%%%%%%%%%%%%%%%%%%%%%%%%%%%%%%%%%%%%%%%%%%%%%%%%%%%%
\begin{table*}[!t]
\centering
\caption{Dataset statistics. RecipeEmission has a comparable sparsity to the baseline datasets. The reduced number of users and items is a result of the filtration steps to improve the efficiency in finding the CO$_2$-eq of the ingredients. We also see a high average rating in both RecipeEmission and Food.com, which needs to be accounted for when comparing the algorithms.}
\label{tab:dataset-comparison}
\resizebox{\textwidth}{!}{%
\begin{tabular}{l c c c c c}
\hline\hline
                                                                             & \textbf{RecipeEmission} & Food.com & Movielens100k~\cite{harper2015movielens} & Movielens1M~\cite{harper2015movielens} & Book-Crossing~\cite{ziegler2005improving} \\ 
                                                                          \rowcolor{gray!50}
\#users                                                                      & 32.090           & 226.570   & 943           & 6.040        & 105.283        \\ 
\#items                                                                      & 5.595            & 231.637   & 1.682          & 3.706        & 340.556        \\ 
\rowcolor{gray!50}
\#interactions                                                               & 247.219          & 1.132.367  & 100.000        & 1.000.209     & 1.149.780       \\ 
sparsity                                                                     & 99.86\%         & 99.998\% & 93.70\%       & 95.53\%     & 99.997\%      \\ 
\rowcolor{gray!50}
rating values                                                               & \{0,\ldots, 5\}               & \{0,\ldots, 5\}         & \{1,\ldots, 5\}              & \{1,\ldots, 5\}            & \{0,\ldots, 10\}             \\ 
average rating (std.)                                                               & 4.57 ($\pm 1.054$)           & 4.41 ($\pm 1.26$)    & 3.53 ($\pm 1.13$)          & 3.58 ($\pm 1.12$)       & 2.87 ($\pm 3.85$)         \\ 
\rowcolor{gray!50}
med. interaction per item $\backslash$ user & 27$\backslash$3              & 2 $\backslash$ 1       & 27 $\backslash$ 65            & 124 $\backslash$ 96         & 1$\backslash$1 \\
\rowcolor{gray!50}
%\begin{tabular}[c]{@{}l@{}}median interactions\\ per item $\backslash$ user\end{tabular} & 27$\backslash$3              & 2 $\backslash$ 1       & 27 $\backslash$ 65            & 124 $\backslash$ 96         & 1$\backslash$1              \\ 
\hline\hline
%\begin{tabular}[c]{@{}l@{}}median interactions \\ per user\end{tabular}      & 3               & 1        & 65            & 96          & 1             \\ \hline\hline
\end{tabular}}
\label{tab_stat}
\end{table*}
%%%%%%%%%%%%%%%%%%%%%%%%%%%%%%%%%%%%%%%%%%%%%%%%%%%%%%%%%%%%%%%%%%%%%%%%%%%%%%%%%%%%%
%%%%%%%%%%%%%%%%%%%%%%%%%%%%%%%%%%%%%%%%%%%%%%%%%%%%%%%%%%%%%%%%%%%%%%%%%%%%%%%%%%%%%

Table~\ref{tab_stat} contrasts the statistics of these datasets. RecipeEmission has fewer users, items, and interactions than Food.com. This is the result of the performed filtration steps to ease the mining the CO$_2$-eq, which also remove noisy/spurious interactions. That is, many users have a few interactions with the items, as shown also in the last row of Table~\ref{tab_stat}. Hence, we selected only recipes with at least $20$ ratings while still preserving the long-tail distribution and a sparsity greater than 99.8\%, in order to retain comparability with conventional RecSys datasets.

RecipeEmission has an average rating of $4.57$, which is close to its maximum of five. This is inherited from the ratings present in the original Food.com dataset. To illustrate the latter, Fig.~\ref{fig:dataset_props} (c) shows that the rating distributions of these two datasets and those of the baselines are similar and skewed towards higher values. Consequently, this skewness will bias the predicted recommendations towards high values. That is, the predicted ratings will have a lower spread which may lead to similar accuracies when comparing different RecSys algorithms. Thus, a relative comparison w.r.t. a baseline algorithm is more insightful to draw conclusions than focusing on absolute numbers. We shall elaborate on the latter in Sec.~\ref{sec:results}.

\begin{figure}[!tp]
\centering
\begin{subfigure}{.4\textwidth}
  \includegraphics[width=1\linewidth]{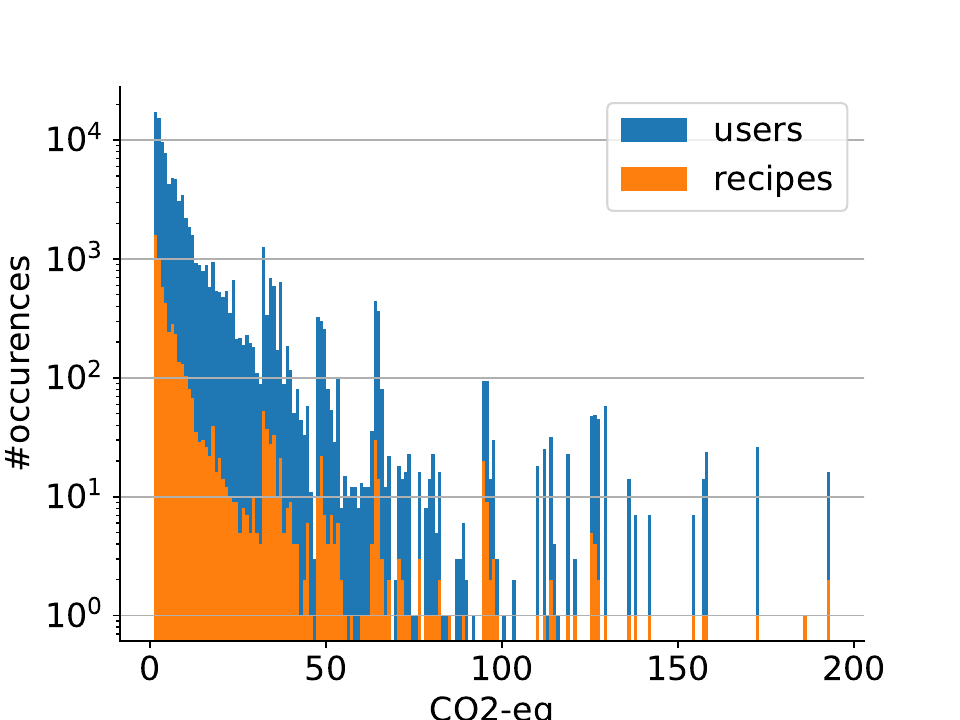}
  \caption{}
  \label{fig:user_item_co2}
\end{subfigure}%
\centering
\begin{subfigure}{.4\textwidth}
  \centering
  \includegraphics[width=1\linewidth]{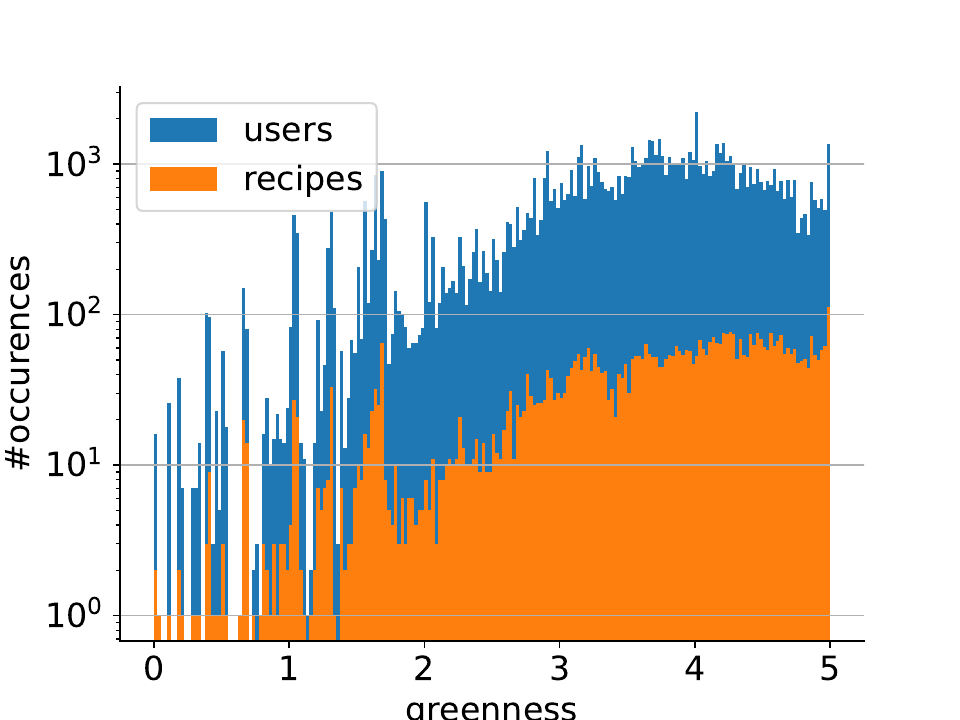}
  \caption{}
  \label{fig:user_item_greenness}
\end{subfigure}%
\vspace{\floatsep}
\begin{subfigure}{.4\textwidth}
  \centering
  \includegraphics[width=1\linewidth]{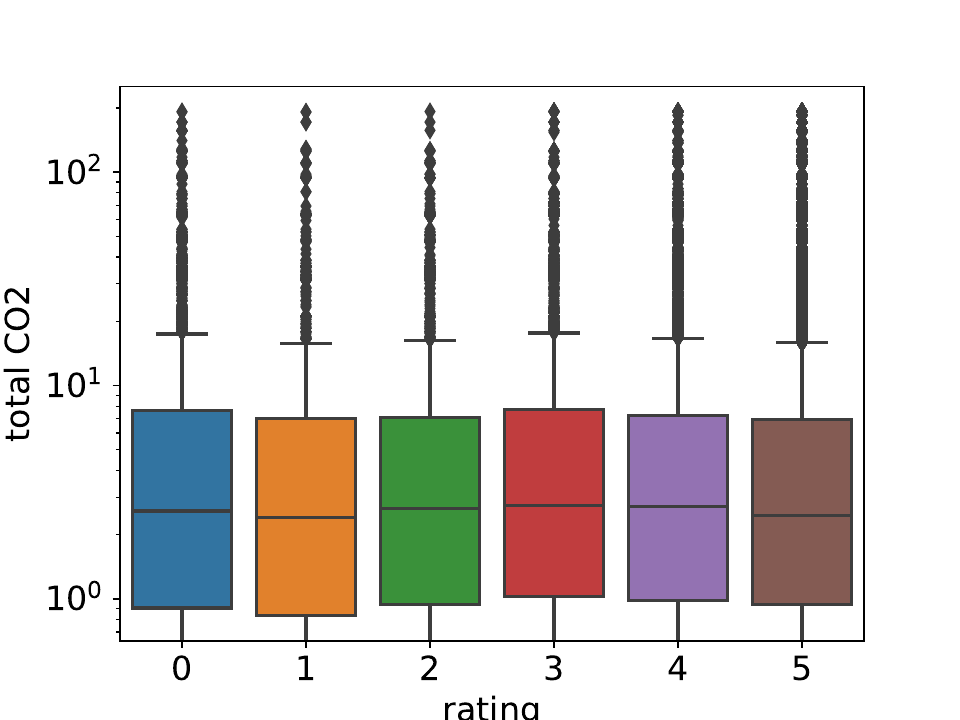}
  \caption{}
  \label{fig:emissions_per_rating}
\end{subfigure}%
\begin{subfigure}{.4\textwidth}
  \centering
  \includegraphics[width=1\linewidth]{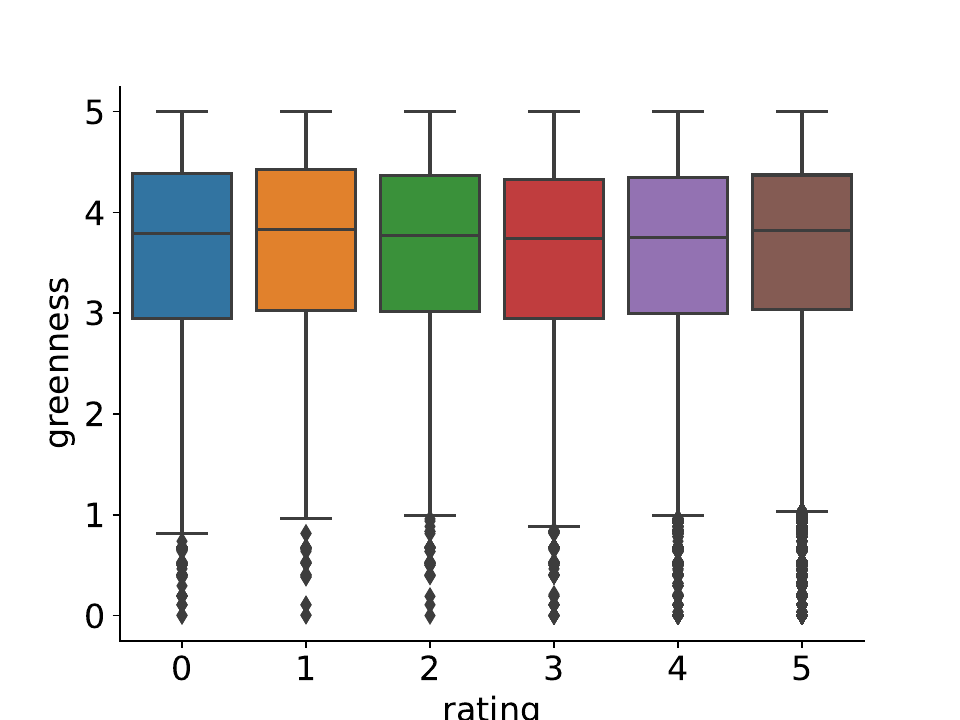}
  \caption{}
  \label{fig:greenness_per_rating}
\end{subfigure}%
\caption{Emission distributions of RecipeEmission dataset. \textbf{(a)} CO$_2$ values of users and items. The high CO$_2$-eq outliers are again the result of recipes with large quantities of high-CO$_2$ products. \textbf{(b)} greenness distributions of RecipeEmissions dataset \textbf{(c)} Distribution of emissions per rating. The large number of outliers is caused by the long-tail distribution of CO$_2$ values of recipes. \textbf{(d)} Greenness distributions over the rating values. There is no apparent distribution shift to the rating, indicating that users did not account for the item greenness when rating them.}
\label{fig:co_rank_rating}
\end{figure}

%
% \smallskip
%\subsection{CO$_2$-eq distribution} 

\smallskip
\noindent\textbf{CO$_2$-eq distribution.}
Fig.~\ref{fig:user_item_co2} shows the CO$_2$-eq distributions of the recipes and users at the interval $[0, 200]$ Kg. Most recipes have a small footprint, and most users interact with recipes with a lower footprint. The decay of these curves indicates a long-tail CO$_2$-eq distribution similar to that of ratings. High CO$_2$-eqs are typically associated with items containing large quantities of high-CO$_2$ ingredients, such as meat products. Most recipes do not contain such high quantities, leading to a low CO$_2$-eq. Thus, the greenness of different recommender systems in absolute terms is unlikely to differ much. 
Analogously, Fig. \ref{fig:user_item_greenness} shows the greenness distribution for the recipes and users. The recipes are skewed towards greener values and, similarly, users interact mostly with greener items. 

Fig. \ref{fig:emissions_per_rating} shows the recipe CO$_2$-eq distribution w.r.t. the ratings. We observe no apparent difference between the distributions, indicating the recipes' carbon footprint plays a minor role in the ratings. 
This is a satisfactory observation, as the distribution of CO$_2$-eq values with respect to the distribution of rating values is unbiased, ensuring it does not interfere with our experimental analysis. For instance, if items with higher CO$_2$-eq values tended to receive higher ratings (e.g., 5 stars), the RecSys model could suffer from a positivity bias~\cite{huang2024going}, leading it to over-recommend these high-rated items. Such a bias would not only skew recommendation quality but also compromise the accuracy of our sustainability analysis by favoring certain groups of items. However, as shown in Fig.~\ref{fig:emissions_per_rating}, our data does not exhibit this correlation, allowing us to conduct both recommendation and greenness analysis without concern for bias related to CO$_2$-eq values. Lastly, in Fig. \ref{fig:greenness_per_rating}, the greenness distribution w.r.t. the ratings is shown. Similarly to the CO$_2$-eq, there is no strong difference between greenness values for different ratings, indicating that users did not consider greenness when rating items.

\subsection{Methods for Building the Dataset}\label{subsec:buildData}

We build the RecipeEmission dataset from the Food.com (the original) dataset present in \cite{majumder2019generating} by adding the CO$_2$-eq and greenness to the different recipe items. This dataset contains users interacting with recipe items via explicit ratings and textual reviews. We break each recipe down into its ingredients (present in the original dataset) and the respective quantities (not present in the original dataset). Upon scraping the ingredient quantities and the CO$_2$-eq, we compute the recipe CO$_2$-eq as a weighted sum of those of the ingredients. This task presents the following intertwined challenges: 
\begin{itemize}
\item[--] \emph{high interaction sparsity:} the original dataset is sparse ($99.998\%$) and often it contains a few interactions per user or item; this leads to some recipes being rated only by one user and to some users having rated only one recipe;
\item[--] \emph{missing ingredient quantities:} quantities of the ingredients are not present in the original dataset and need to be found differently;
\item[--] \emph{non-unified description:} descriptions of the quantities differ (e.g., 1-2 cups, 1.5 grams) and they are often expressed in qualitative terms (1 piece, a pinch);
\item[--] \emph{lack of CO$_2$-eq consensus:} there is no standard and unified repository to find the CO$_2$-eq of the different ingredients (this differs from the country of origin, the technology used to harvest them, transportation to destination, and season); this often requires extensive manual work;
\item[--] \emph{missing CO$_2$-eq values:} CO$_2$-eqs are missing for particular ingredients, leading to recipes having only some ingredients for which the CO$_2$-eq could be found.
\end{itemize}
Here, we present the methods pursued to address these challenges. These methods consider the thin balance between \emph{including as many interactions as possible} and \emph{having reliable CO$_2$-eq measures} for the items. Our focal principle is to preserve the key properties (see Fig.~\ref{fig:dataset_props}) and statistics (see Table~\ref{tab_stat}) of the original dataset to ensure that they remain comparable to baselines. The methods are composed of five steps: $i)$ pre-filtering the data; $ii)$ quantifying the ingredients in the recipes; $iii)$ quantifying the CO$_2$-eq; $iv)$ quality assessment; $v)$ quantizing the CO$_2$-eq. Fig.~\ref{fig:dataset_building} provides a high level overview of the process of building the dataset.

% \begin{figure}
%   \includegraphics[width=\textwidth]{Figures/dataset_building.pdf}
%   \caption{Overview of the steps taken to build the dataset. First of all, pre-filtering is performed, followed by ingredient and CO$_2$-eq quantification. Finally, the CO$_2$-eq values are transformed to a greenness scale.}
%   \label{fig:dataset_building}
% \end{figure}
% \documentclass[tikz,border=9pt]{standalone}
\usetikzlibrary{shapes, arrows.meta, positioning}
% \begin{document}
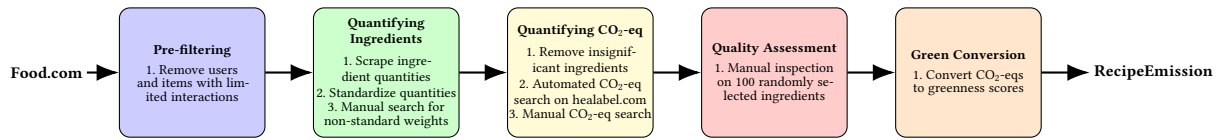
\begin{figure}
% \begin{minipage}[t]{.1\textwidth}
\begin{tikzpicture}[
    box/.style={
        rectangle, draw, rounded corners, text width=4.5cm, align=center,
        minimum height=4cm, inner sep=0pt
    },
    arrow/.style={
        -{Latex[length=3mm]}, thick
    },
    scale=0.43, transform shape
]
% Define nodes aligned by their top edge using anchor=north
\node[box, anchor=north, fill=blue!20] (prefilter) at (0,0) {%
    \Large\textbf{Pre-filtering} \\[2mm]
    1. Remove users and items with limited interactions
};

\node[box, anchor=north, fill=green!20] (quantIngredients) at ([xshift=6cm] prefilter.north) {%
    \Large\textbf{Quantifying Ingredients} \\[2mm]
    1. Scrape ingredient quantities\\
    2. Standardize quantities\\
    3. Manual search for non-standard weights
};

\node[box, anchor=north, fill=yellow!20] (quantCO2) at ([xshift=6cm] quantIngredients.north) {%
    \Large\textbf{Quantifying CO$_2$-eq} \\[2mm]
    1. Remove insignificant ingredients\\
    2. Automated CO$_2$-eq search on healabel.com\\
    3. Manual CO$_2$-eq search %for remaining ingredients
};

\node[box, anchor=north, fill=red!20] (quality) at ([xshift=6cm] quantCO2.north) {%
    \Large\textbf{Quality Assessment} \\[2mm]
    1. Manual inspection on 100 randomly selected ingredients
};

\node[box, anchor=north, fill=orange!20] (greenConversion) at ([xshift=6cm] quality.north) {%
    \Large\textbf{Green Conversion} \\[2mm]
    1. Convert CO$_2$-eqs to greenness scores
};

% Incoming arrow labeled "Food.com" into the Pre-filtering box
\draw[arrow] ([xshift=-1cm] prefilter.west) -- (prefilter.west)
    node[pos=0, left] {\LARGE\textbf{Food.com}};

% Arrows connecting the boxes
\draw[arrow] (prefilter.east) -- (quantIngredients.west);
\draw[arrow] (quantIngredients.east) -- (quantCO2.west);
\draw[arrow] (quantCO2.east) -- (quality.west);
\draw[arrow] (quality.east) -- (greenConversion.west);

% Outgoing arrow labeled "RecipeEmission" from Green Conversion (no box)
\draw[arrow] (greenConversion.east) -- ++(1.5cm,0)
    node[pos=1, right] {\LARGE\textbf{RecipeEmission}};
\end{tikzpicture}
\caption{Overview of the steps taken to build the dataset. First of all, pre-filtering is performed, followed by ingredient and CO$_2$-eq quantification. Finally, the CO$_2$-eq values are transformed to a greenness scale.}
  \label{fig:dataset_building}
% \end{document}
\end{figure}
% \end{minipage}

%%%% PREFILTERING
\smallskip
\noindent\textbf{Pre-filtering.} The long-tail distribution of the Food.com dataset (see Fig.~\ref{fig:dataset_props}) implies that we need to find the CO$_2$-eq for some items that play a minor role in assessing the overall performance of a RecSys algorithm\footnote{These items would, however, be relevant if we were interested in addressing the (strict) cold start problem for RecSys. But since our focus is on assessing the \emph{greenness} of RecSys algorithms as a whole, we can ignore these items. Typically, these interactions are treated as spurious or erroneous and are removed.}. To reduce the number of recipes for which the CO$_2$-eq has to be found, we filtered out the items and users with a few interactions as commonly done in RecSys datasets --e.g., MovieLens filters on the minimum number of interactions per user~\cite{harper2015movielens}. Specifically:
%%%%%%%
\begin{enumerate}[label=\arabic*)]
%%%%%%%%%%%%%%%%%%%%%%%%%%%%%%%%%%%%%%%%%%%%%
\item We removed all recipes having less than $20$ ratings.
%%%%%%%%%%%%%%%%%%%%%%%%%%%%%%%%%%%%%%%%%%%%%
\item From the remaining recipes, we removed those for which not all ingredients could be found.
%%%%%%%%%%%%%%%%%%%%%%%%%%%%%%%%%%%%%%%%%%%%%
\item Further, we removed all users that had no interaction with the remaining recipes.
%%%%%%%%%%%%%%%%%%%%%%%%%%%%%%%%%%%%%%%%%%%%%
\item From the recipes with more than $20$ interactions, we removed those users that have rated only that recipe, while still maintaining at least $20$ interactions per recipe.
%%%%%%%%%%%%%%%%%%%%%%%%%%%%%%%%%%%%%%%%%%%%%
\item {Finally, we removed those recipes for which all users have rated less than 20 recipes on average. That is, for each recipe, we extracted the user IDs of users who have interacted with the recipe and measured their total number of interactions. Then, we removed those recipe items for which the respective users rated less than 20 items. The latter helps us remove recipes that occur only in sparse parts of the dataset and tackle the cold start issues.} %\red{@Raoul: check what I wrote here based on our long-time ago discussion. I recall though there you were talking about "averages". Could you explain it here and update this part accordingly?} 
This led to the final RecipeEmissions dataset reported in Table~\ref{tab_stat}.
%
%\blue{Finally, we also removed recipes for which all users rated less than 20 recipes on average.}\red{EI: I don't get this! Does it mean we removed those recipes that had again less than $20$ ratings?} \green{Explanation: first we removed all recipes with less than 20 ratings (so less than 20 users). Later on, we removed also recipes for which its \emph{users} had less than 20 ratings \emph{on average}. In other words, we calculate the average number of interactions among users of every recipe and we remove all recipes for which this is less than 20. We discussed this a long time ago, and the main reason we did this, was to remove recipes that only occured in very sparse parts of the dataset. We figured this would be likely to not change the results much, while at the same time it would cost much to find their CO2.}\red{EI: Still unclear to me! We need to talk about this; so you can explain it to me in 5 min.} This led to the final RecipeEmissions dataset reported in Table~\ref{tab_stat}.
\end{enumerate}

%%%% QUANTIFY INGREDIENTS
\smallskip
\noindent\textbf{Quantifying ingredients.} The original dataset from \cite{majumder2019generating} does not contain the ingredient quantities, which we got from Food.com:
%%%%%%%%%%%%%%%
% ENUMERATE
\begin{enumerate}[label=\arabic*)]
%%%%%%%%%%%%%%%%%%%%%%%%%%%%%%%%%%%%%%%%%%%%%
\item \emph{Scrape:} We scraped Food.com using the URL:
\begin{center}
\emph{www.food.com/[tag]/[recipe name]-[recipe id]}.\smallskip\\ 
\end{center}
Here \emph{[tag]} is \emph{recipe} or \emph{about} depending on the recipe. The \emph{[recipe name]} is comprised of the decapitalized version of the recipe name with spaces replaced by hyphens (-). The \emph{[recipe id]} is also available in the original dataset.

%%%%%%%%%%%%%%%%%%%%%%%%
%%%%%% TABLE
\definecolor{Gray}{gray}{0.9}
\begin{table}[t]
\caption{Ingredient categories, their representations, and their corresponding weights conversion from pint, cup, teaspoon, and tablespoon to grams. "na." implies that measuring approach was not present for the specific category. The "pinch" metric was used for spices and salt and we choose salt as the representative ingredient with a conversion rate of 0.36 grams.}
\label{tab:my-table-conv}
\begin{tabular}{lccccc}
\hline\hline
\rowcolor{Gray}
{\textbf{Category}}     & \textbf{Represented by} & \textbf{Pint-to-g} & \textbf{Cup-to-g} & \textbf{Teaspoon-to-g} & \textbf{Tablespoon-to-g} \\
{liquids}               & water                   & 473.18   & 236.59   & 4.93 & 14.79         \\
\rowcolor{Gray}
{sugars and sweeteners} & raw sugar               & 414.02   & 227.36  & 4.74  & 14.21        \\
{flours}                & all-purpose flower      & 250.31   & 125.16 & 2.61 & 7.82         \\
\rowcolor{Gray}
{oils and fats}         & olive oil               & 434.38    & 217.66   & 4.53  & 13.6       \\
{spices}                & cinnamon                & 264.98   & 132.49  & 2.76  & 8.28         \\
\rowcolor{Gray}
{nuts}                  & almonds                 & 217.66     & 108.83  & na.  & na.       \\
{fruits and vegetables} & onion                   & 104.10  & 150    & na.  &na.      \\
\hline\hline
\end{tabular}
\end{table}
%%%%%%%%%%%%%%%%%%%%%%%%

%
All ingredients on Food.com have unique identifiers, but those do not always correspond to the identifiers in the original dataset. However, the order of ingredients on Food.com follows that in the lists. We contrasted the list lengths and the ingredient names to find the corresponding entities, and we also checked this manually for $50$ random recipes. Thus, we scraped the ingredient quantities and respective units in their default order from Food.com. \smallskip

For 797 recipes, the ingredient list on Food.com contained more ingredients than the list in the original dataset. These additional ingredients formed alternatives or minor additions that do not change the recipe. We checked if the names of the ingredients in the original dataset were contained in the names of scraped ingredients. If so, we matched the two and retrieved the respective quantities; otherwise, we considered the scraped ingredient not to be present in the original dataset and removed the ingredient.
%%%%%%%%%%%%%%%%%%%%%%%%%%%%%%%%%%%%%%%%%%%%%
\item \emph{Standardization:} The quantities on Food.com are expressed in different forms (e.g., range 1-2 cups; decimal format 2.5 cups; decimal fractions 2 1/2 cups) and in different units (grams, ounce, pinch). We standardized the quantities into a decimal format and in grams to allow for an easier CO$_2$-eq conversion, which is expressed in Kg. The conversion rates are reported in Table~\ref{tab:my-table-conv}.
%%%%%%%%%%%%%%%%%%%%%%%%%%%%%%%%%%%%%%%%%%%%%
\item \emph{Manual completion:} To estimate the weight of ingredient instances in non-standard quantities (e.g., "two chicken breasts," "a bunch of cabbage," or "a handful of cabbage"), we engaged four student annotators from diverse demographic and cultural backgrounds. Each annotator was tasked with determining the weight in grams of various ingredient-measure combinations sampled from our dataset. They were instructed to search for the ingredient and measure combination on Google.com and record the first relevant result they found. Additionally, they were asked to provide their source and rate their confidence in the accuracy of their findings on a scale from 1 to 10 to support the verification process. To enhance efficiency and consistency, we provided custom software that allowed annotators to input both their results and confidence levels directly. The software randomly assigned ingredient-measure combinations, and each annotator processed half of the dataset, ensuring all ingredients were annotated multiple times.

Each of the 4,010 non-standard ingredient instances was annotated. Estimates with confidence ratings below 5 were re-examined, leading to 73 re-annotations. To assess inter-rater reliability, we calculated the mean absolute difference between annotations from different raters, which revealed a disagreement of a mere 2\%, indicating a high level of reliability.
%\textcolor{red}{An annotator estimated the weights for each of the $4.010$ ingredient instances that were in non-standard quantities and with repetition (e.g., two chicken breasts, a bunch of cabbage, or a handful of cabbage) by assigning it to the first search result on Google.com. 
%We found the quantities for \blue{$4.010$} ingredients which were in non-standard quantities (e.g., two chicken breasts or one bell pepper). A manual annotator estimated the weights of each ingredient by assigning it to the first search result on Google.com. 
%Annotators also provided the source and a numerical value (1-10) of how certain they were about the correctness of the annotation. Annotations with certainties below $5$ were re-inspected, resulting in $73$ re-annotations. \textcolor{blue}{Moreover, as a measure of the inter-rater reliability, we calculated the mean absolute difference between the annotations of the different raters. This showed a mere $2\%$ disagreement between annotators, indicating a high reliability of annotations}}
\end{enumerate}

%%%%%QUANTIFYING-CO2
\smallskip
\noindent\textbf{Quantifying CO$_2$-eq.} Aided by the list of ingredients, the respective quantities, and the standardized units, we found the CO$_2$-eq of the $2735$ unique ingredients as follows: %\blue{$2735$}\footnote{\red{EI: above we have 4010 ingredients in non-standard quantities but we also had other ingredients, which if I'm not wrong were quite a few? Can we give that all number somewhere? But where we say we found the co2-eq for 2735 ingredients. So, one would ask what happened to the remaining ones. We shall mention somewhere this!} \green{I am not sure I was clear here: there is a number of ingredients, i.e. 2735, but the ingredients are used in different instances. So, we have the ingredient cabbage, for example. Cabbage has a fixed greenness value per kg. But, cabbage is used in different instances (so, a pack of cabbage, a bunch of cabbage, and a handful of cabbage. That is why there are more instances (i.e... 4010 non-standard ones) than ingredients.)}} ingredients as follows:
%%%%%%%%%%%%%%%
% ENUMERATE
\begin{enumerate}[label=\arabic*)]
%%%%%%%%%%%%%%%%%%%%%%%%%%%%%%%%%%%%%%%%%%%%%
\item \emph{Ingredient filtration:} First, we removed $705$ ingredients present in quantities smaller than $50$ grams. These ingredients are in small relative quantities or have a low CO$_2$-eq, such as spices and flavorings \cite{marie_2022, marie_2022_2}. The threshold of $50$ grams is chosen as that of having a low impact on the whole recipe CO$_2$-eq, since in many recipes, the overall CO$_2$-eq is governed by a few ingredients present in large quantities and with a large carbon footprint (e.g., meat products).%, the threshold of $50$ grams is qualitatively chosen as that of having a relatively low impact on the whole recipe CO$_2$-eq.
%%%%%%%%%%%%%%%%%%%%%%%%%%%%%%%%%%%%%%%%%%%%%
\item \emph{Automated search:} We searched the CO$_2$-eq of all ingredients on healabel.com. This source, which provides the most extensive list of ingredients with CO$_2$-eq data, aggregates information from books, journals, and validated online articles, including the FDA's Food Resources for Consumers, the World Health Organization, and the US Department of Agriculture. Having an extensive collection of CO$_2$-eq, this website also allowed a consistent measure across the different items in the dataset. We extracted the CO$_2$-eq via the ingredient names using the URL:
\begin{center}
\emph{https://healabel.com/ [first letter ingredient] / [ingredient name]}.
\end{center}
Multiple word names are separated with a hyphen (-). We searched ingredients both in their singular and plural forms (e.g., nut and nuts) as both were present. Several ingredients are variations of others (e.g., red pepper, green pepper) and have comparable CO$_2$-eq. So, we removed the adjectives and searched their stem form (e.g., pepper). For some composite words, the adjective represents the main ingredient (e.g., meat substitute). {However, for such ingredients, it is inaccurate to use the adjective to find the CO$_2$-eq. For example, a meat steak differs from a meat sauce. Thus, we found the CO$_2$-eqs of these ingredients manually.}
%%%%%%%%%%%%%%%%%%%%%%%%%%%%%%%%%%%%%%%%%%%%%
\item \emph{Manual search:} We searched the CO$_2$-eq first on healabel.com to standardize the data as much as possible and then in literature. The latter is achieved by searching on Google Scholar using the query \emph{ [ingredient name] CO$_2$ footprint} and retrieving the CO$_2$-eq from the most cited paper. The papers used for this step are \cite{yuttitham2011carbon, sevenster2008sustainable, dalla2017environmental, fantozzi2019life, parajuli2021cradle, taylor2014greenhouse, vasilaki2016water, pelletier2013carbon, espinoza2011carbon}.
%%%%%%%%%%%%%%%%%%%%%%%%%%%%%%%%%%%%%%%%%%%%%
\end{enumerate}

%%%%QUALITY CHECK
\smallskip
\noindent\textbf{Quality assessment.} To evaluate the quality of the dataset, $100$ random ingredients were selected, and the accuracy of their retrieved CO$_2$-eq was manually inspected. Due to the complexity of this process, we evaluated the similarity between the ingredient name and the corresponding name used to retrieve the CO$_2$-eq. For example, retrieving the exact CO$_2$-eq value for the ingredient "1 cup of shredded bell pepper" is a complex task, or in some cases, even impossible. Hence, we tackled this issue by retrieving the CO$_2$-eq value for the most similar ingredient to "1 cup of shredded bell pepper" with a known CO$_2$-eq. In this case, the ingredient "bell pepper" would be the best option. The resulting accuracy of this evaluation was $92\%$, indicating that for $92\%$ of the ingredients, we could accurately retrieve the CO$_2$-eq value. We deemed this accuracy satisfactory as we are not interested in the precise CO$_2$-eq but in relative quantification. A more precise quantification should start by standardizing the quantities of the different ingredients, their densities, the dynamics of CO$_2$-eq (season, distance from the user), and the sources from which the CO$_2$-eq is retrieved. This by itself requires extensive separate research.\footnote{The complexity arises from the lack of global standardization in CO$_2$-eq and weight conversions for ingredients. For example, the conversion rate for \emph{a portion of beef} differs substantially between Texas and the Netherlands. Factors such as the breed and age of the animal further complicate this. Therefore, constructing a unified dataset for detailed CO$_2$-eq analysis would require extensive interdisciplinary research, which goes beyond the scope of a single paper.}

%%%%%GREEN CONVERSION
\smallskip
\noindent\textbf{Green conversion.} Each of the above steps introduces some uncertainty to the precise CO$_2$-eq of whole recipes. Factors contributing particularly to this uncertainty include the precise weight calculation of the ingredients, manual annotations, removal of the ingredients with small quantities, variations in ingredient names, accuracy of CO$_2$-eq retrieval, and the sampling used in quality assessment. To reduce the effect of this uncertainty and to be consistent with the interval of rating values, we transformed the CO$_2$-eq into \emph{greenness} values at the interval $[0, 5]$, where a low value corresponds to a less green item (high CO$_2$-eq) and a high value to a greener item (low CO$_2$-eq).
%
%The CO$_2$-eq in Kgs of the retried recipes have an uncertainty around the indicated value because of the tedious task of finding a unified way of quantifying the CO$_2$-eq. To reduce the effect of this uncertainty, we quantized the CO$_2$-eq into a discrete interval as done for explicit ratings. More precisely, we quantized the CO$_2$-eq into \emph{greenness} values in the interval $\{0,\ldots, 5\}$, where a low value indicates a less green item (high CO$_2$-eq) and a low value a greener item (low CO$_2$-eq). This greenness conversion is done so that the interval matches that of the ratings and also has a similar ascending order for a more "useful item" (high rating, high greenness). 
%
Specifically, denoting the CO$_2$-eq of item $i$ by $c_i$, the greenness $g_i$ is defined as
\begin{equation}\label{eq_greenness}
g_i = \ccalQ\left[\log\left(1+ \frac{1}{c_i}\right)\right]
\end{equation}
where the inverse ensures that items of low CO$_2$-eq are greener and vice versa; the logarithm amortizes the exponential decay spreading of the CO$_2$-eq; and $\ccalQ[\cdot]$ is a function that projects the greenness values into $[0, 5]$ interval, e.g., values higher than 5 are mapped to 5. We added a one within the logarithm to have only positive values inspired by the tf-idf weighting \cite{salton1988term}. This greenness conversion matches that of the ratings and has a similar ascending order for a more "useful item" (high rating, high greenness).
\section{Benchmarking and Trade-offs}\label{sec:results}

The goal of this section is twofold. First, we benchmark different recommender system algorithms to answer RQ2 in Sec.~\ref{subsec:benchmarkResults}. Second, we study the accuracy-greenness trade-off to answer RQ3 in Sec.~\ref{subsec:accGreen_trade}.

\subsection{Experimental Setup}

\smallskip
\noindent\textbf{Algorithms.} We consider the following nine algorithms: i) \emph{random predictions} and ii) \emph{global mean} to act as a baseline for relative comparisons of; iii) \emph{Item nearest neighbor (ItemNN)}~\cite{sarwar2001item}; iv) \emph{User nearest neighbor (UserNN)}~\cite{resnick1994grouplens}; v) \emph{Singular value decomposition (SVD)}~\cite{sarwar2002incremental}; vi) \emph{SVD++} \cite{koren2008factorization} that enhances SVD by accounting for implicit ratings; vii) \emph{CoClustering} \cite{george2005scalable} that uses k-means to assign similar users/items in the same cluster and uses the mean cluster rating for prediction; viii) \emph{sparse linear method (Slim)} \cite{ning2011slim}, an ElasticNet-based neighboring approach; {ix) \emph{graph neural network {GNN}} \cite{berg2017graph} that uses graph convolutions to complete the rating matrix. Note that none of these recommendation algorithms incorporates greenness information. %during the training process or in building the model.

We wanted to establish a baseline (random and global mean) that would allow a relative comparison of the conventional neighboring-based (ItemNN and UserNN), matrix factorization (SVD and SVD++) and clustering-based (CoClustering) approaches. We included Slim as an additional neighboring-based approach since it has shown a consistent performance in RecSys tasks\footnote{{https://www.amazon.science/latest-news/amazon-scholar-george-karypis-receives-icdm-10-year-highest-impact-award}}. GNNs have been considered as a more representative alternative to neighboring-based approaches since they can capture information from multi-hop neighbors. This set of algorithms has been kept simple for three reasons. First, they are the backbone of a myriad of more sophisticated solutions; hence, analyzing them will serve as a baseline for the follow-up alternatives. Second, they often achieve a comparable performance w.r.t. more complex solutions \cite{ferrari2019we}. And third, more involved deep learning solutions often require extensive hyperparameter tuning that is beyond our scope. We detail their inner-working mechanism in Appendix~\ref{appendix}.

\smallskip
\noindent\textbf{Metrics.} We rank the items based on the estimated ratings and measure the list's accuracy and greenness. We measure the accuracy of a list of length $k$ via the normalized discounted cumulative gain at $k$ (NDCG$@k$) that considers higher-positioned items more relevant. We measure the greenness of a list by following a similar principle via a metric that attains a high value if the higher-positioned items are greener. Denoting the greenness of item $i$ by $g_i$ and its ranking in the list by $v_i$, we define the green DCG$@k$ (GDCG$@k$) as:
\begin{equation}\label{eq:GDCG}
\text{GDCG@}k=\frac{1}{|U|} \sum_{u=1}^{|U|} \sum_{i \in \ccalI_{u}, v_{i} \leq k} \frac{2^{g_{i}}-1}{\log _{2}\left(v_{i}+1\right)}
\end{equation}
where $U$ is the set of users and $\ccalI_u$ the set of items user $u$ has interacted with.
%
%\red{EI:
%\begin{enumerate}
%\item (Previous discussion) Elvin: why there is a double sum; typically it is only one sum only \green{Raoul: In the original NDCG formula, this is also done like this I think, so I did it in the same way to keep the similarities.};
%\begin{itemize}
%\item (Cont...) can you give me a reference for that to double-check; add it here! \cite{aggarwal2016recommender} (The DCG is in chap. 7.5.3)
%\end{itemize} 
%\end{enumerate}
%}

\begin{figure}
  \includegraphics[width=\textwidth]{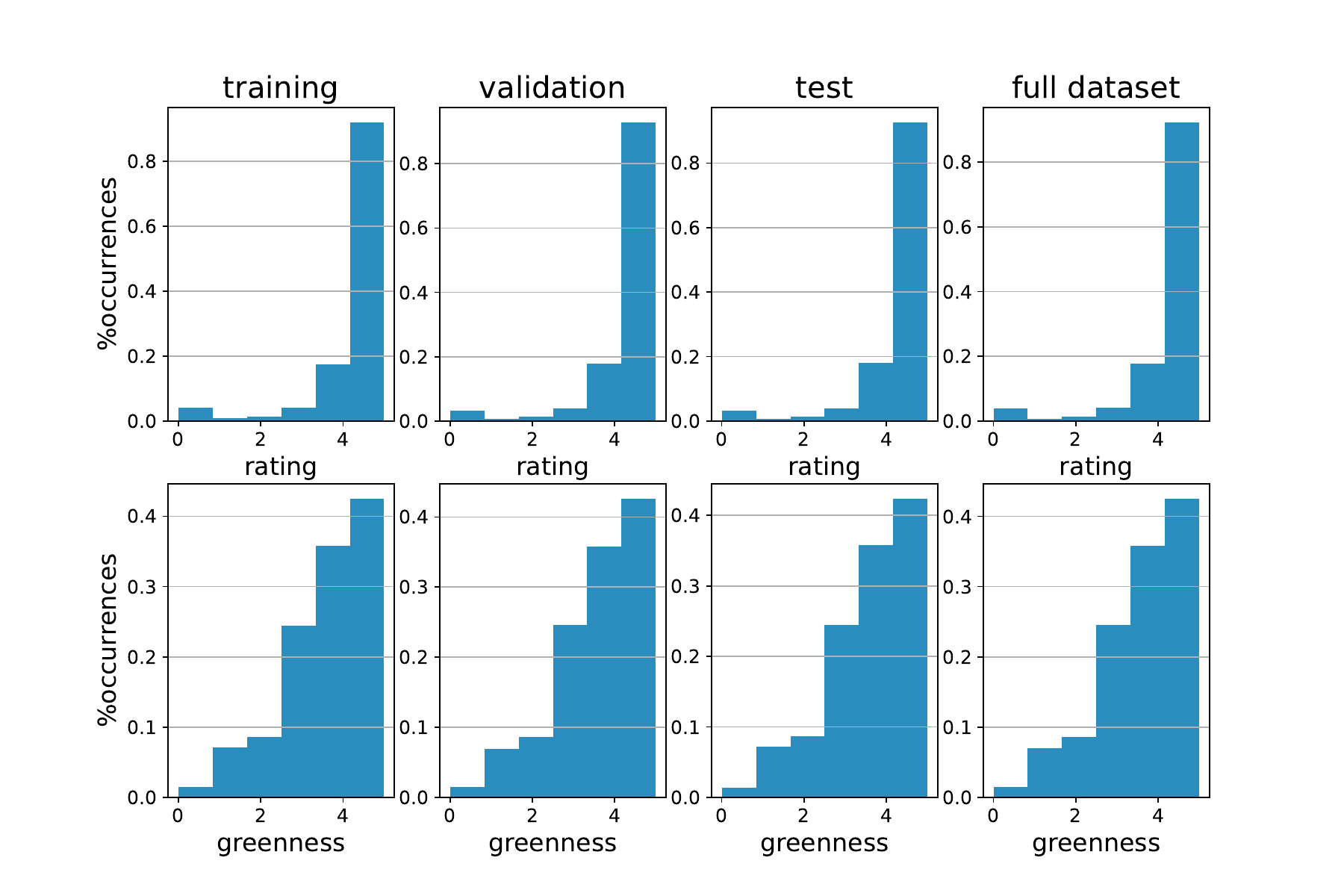}
  \caption{Rating and greenness distributions of the training, validation and testing splits and the full datasets. Sparsities are 99.9\%, 99.9\%, 99.9\% and 99.9\% respectively.}
  \label{fig:recipe_entry}
\end{figure} 

The numerator increases exponentially with the item greenness (rather than item relevance as in DCG) so that highly-ranked green items yield a high GDCG$@k$. The normalized $\text{GDCG@}k$ ($\text{GNDCG@}k$) is obtained by dividing \eqref{eq:GDCG} by the ideal $\text{GDCG@}k$, where the latter is computed by ranking items in the list based on their actual greenness. The $\text{GNDCG@}k$ attains values in the interval $[0,1]$ where a higher value indicates a greener list and vice versa. The greenness of an item $i$, $g_i$, is obtained by mapping CO$_2$-eq values into real values into $[0, 5]$. We refer to Sec.~\ref{sec:data} for more details on the greenness computation. %\textcolor{magenta}{EI: if we provide the greenness score earlier in the paper, we can refer to it here. This will help the flow and we don't need this text in blue anymore.}

\smallskip
\noindent\textbf{Data split.} We split the data with a $6/2/2$ ratio into training set (148,512 interactions), validation set (49,444 interactions), and test set (49,444 interactions). We considered all users and items to be present in all sets to avoid strict cold start issues, we created the train-test split as follows: we started with an empty test set and a training set containing all interactions. We selected an interaction at random for which both the user and item occurred at least two times in the training set. This interaction was then removed and transferred to the test set. We repeated this process until the test set contained at least $20\%$ of all interactions. Using a similar process, we built the validation set from the remaining data in the training set.
{This ensures that all items in the test set also occur at least once in the training set. Users and items that occur only once in the dataset can, therefore, not occur in the test set. Also, the random picking strategy makes it possible that not all users and items occur in the test set. The distribution of the ratings and greenness of the splits is shown in Fig. \ref{fig:recipe_entry}}.

We also preserve similar greenness distributions among the sets and the whole dataset to avoid distribution biases. %Details of the procedure are reported in the Method Sec.~\ref{subsec:data_split}.
We build five such random splits to measure the performance of the algorithms. When measuring the NDCG$@k$, we further split the test set into batches of $100$ samples. Then, we measure the NDCG$@k$ in each batch and average across all batches. The reason for this is that the test set inherits the rating bias from the RecipeEmission dataset (Table.~\ref{tab:dataset-comparison}, Fig.~\ref{fig:dataset_props} (c), Fig.~\ref{fig:recipe_entry}); i.e., $77\%$ of the test samples have a rating of five.
%i.e., out of the $49.444$ test samples, $38.187$ have a rating of five. 
Consequently, measuring the NDCG$@k$ on the whole test set would typically lead to a value of one and would not discriminate the algorithms in terms of accuracy. Note that the different splits are all ranked and evaluated independently. %The proposed batch division avoids this issue.

\begin{table*}[!t]
\centering
\caption{Average performance of the different recommender systems algorithms for different list length in a ranking-based setting. The standard deviation is approximately $2-5 \times 10^{-3}$ except for the instances marked with $^*$ which is $10^{-2}$.}
\label{tab:dataset-comparison}
\begin{tabular}{l c c c || c c c}
\hline\hline
Algorithm & NDCG@$10$ & NDCG@$20$ & NDCG@$50$ & GNDCG@$10$ & GNDCG@$20$ & GNDCG@$50$\\
\hline
 \rowcolor{gray!50}
Random & 0.86   & 0.86 & 0.86 & 0.47   & 0.51$^*$   & 0.61$^*$  \\ 
Global Mean & 0.87   & 0.86 & 0.86 & 0.46   & 0.51$^*$   & 0.62$^*$  \\ 
\rowcolor{gray!50}
ItemNN & 0.92   & 0.91  & 0.91 & 0.47   & 0.51  & 0.61 \\
UserNN & 0.89 & 0.89    & 0.87  & 0.49    & 0.52  & 0.62 \\
\rowcolor{gray!50}
SVD & 0.97    & 0.95   & 0.93 & 0.49   & 0.53   & 0.63 \\
 SVD++ & 0.96  & 0.95   & 0.93 & 0.49   & 0.52  & 0.62  \\
 \rowcolor{gray!50}
 CoClustering & 0.95   & 0.94   & 0.92 & 0.48$^*$   & 0.53  & 0.62 \\
  Slim & 0.95    & 0.94   & 0.92  & 0.49  & 0.53  & 0.62 \\
  \rowcolor{gray!50}
   GNN & 0.96   & 0.95   & 0.92 & 0.49   & 0.52  & 0.63 \\
\hline\hline
%\begin{tabular}[c]{@{}l@{}}median interactions \\ per user\end{tabular}      & 3               & 1        & 65            & 96          & 1             \\ \hline\hline
\end{tabular}
\label{tab_ratPerf}
\end{table*}

\subsection{Benchmarking}\label{subsec:benchmarkResults}

We optimize the algorithms for accuracy and evaluate their greenness in the test set. %The RMSE is used as a cost function to estimate the algorithm parameters and the lists are built from the highest estimated ratings. 
From the results in Table~{\ref{tab:dataset-comparison}}, we make three key observations.

\smallskip
\noindent\textbf{RecSys algorithms ignore greenness.} Although RecSys algorithms do not utilize the greenness information as input, for all list lengths, the GNDGC of all algorithms is around the middle value of 0.5. This shows that when optimized for accuracy, RecSys algorithms do not prioritize the greener or less green items. The latter is in part reminiscent of the CO2$_2$-eq and greenness distributions in Figs.~\ref{fig:co_rank_rating} (a) and (b). The greenness distribution centers on the neutral value of three, and the ratings have all a similar CO$_2$-eq distribution; see Fig.~\ref{fig:dataset_props} (d). Differently, the NDCG is greater than 0.9. However, we should judge this performance relative to the Global Mean and Random predictions because the dataset has a high-rating bias. All the algorithms improve over these baselines.

%\begin{figure}
%%\begin{figure}{R}{0.375\textwidth}
%\begin{center}
%  \includegraphics[width=.5\linewidth]{Figures/greenness_per_rating.pdf}
%\end{center}
%\caption{Greenness distribution over rating values.}
% \label{fig:greenness_per_rating}
%\end{figure}

\begin{wrapfigure}{r}{0.5\textwidth}
\vskip-.5cm
\begin{minipage}{80mm}
\begin{center}
  \includegraphics[width=\linewidth]{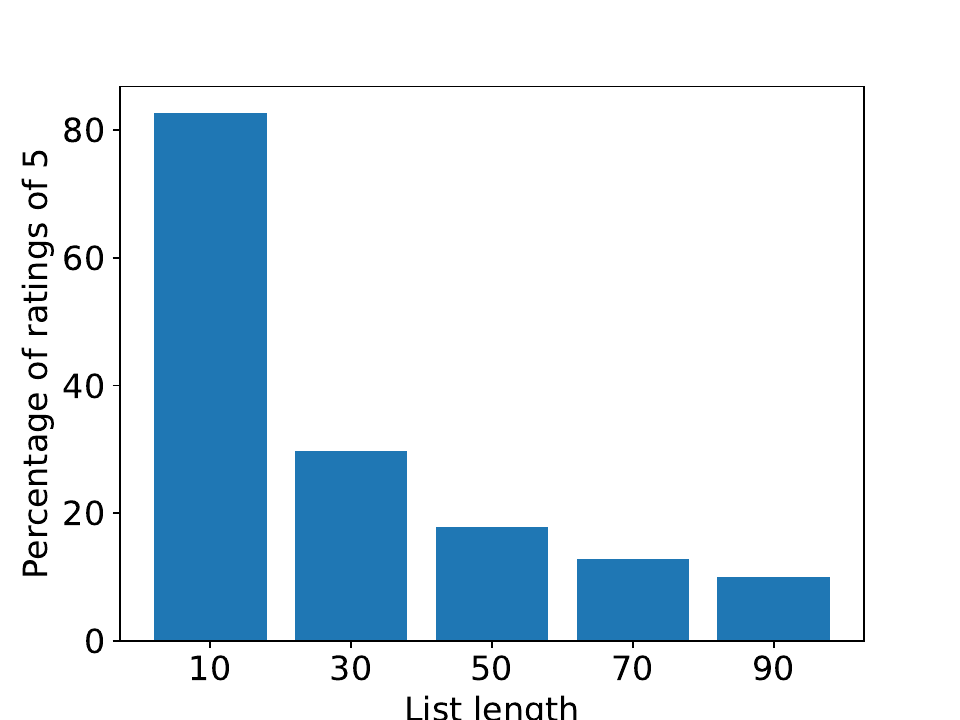}
  \end{center}
    \captionof{figure}{Average percentage of ratings of 5 for different list lengths, resulting from the SVD predictions.}
        \label{fig:percentage_ratings_5}
  \end{minipage}
  \vskip-.25cm
  \end{wrapfigure}

\smallskip
\noindent\textbf{Longer lists are greener but less accurate.} All algorithms attain a higher GNDCG for longer lists. We attribute this to the greenness distribution (Figs.~\ref{fig:co_rank_rating} (a) and (b)) and to the GDCG$@k$ metric (equation~\ref{eq:GDCG}). Regarding the distribution, a larger portion of the recipes have a greenness of over two. Therefore, a longer list has a higher chance of including a greener item. Regarding the metric, the item greenness affects the GDCG$@k$ as $2^{g_i}$; i.e., it rewards rankings with higher-positioned greener items. This approach regards greener items with a lower position in the list as more valuable than non-green items that are slightly higher. Since the items are not ranked based on their greenness, a longer list includes more greener items (see distribution). And since the lower-positioned greener items contribute more than the non-green higher-positioned items, the GDCG, and with that the NGDCG, grows.

Differently, accuracy is affected by the list length. Because of the rating bias in the dataset, in shorter lists, the relative number of items with a true rating of five is higher than that in longer lists; hence, the NDCG reduces; see an example in Fig. \ref{fig:percentage_ratings_5}. We could see this behavior as an accuracy-greenness trade-off w.r.t. the list length.
%%%%%%%%%%%%%%%%
%\begin{figure}
%\begin{figure}{R}{0.375\textwidth}
%\begin{center}
%  \includegraphics[width=.5\linewidth]{Figures/listlength_vs_rating5.pdf}
%\end{center}
%\caption{Average percentage of ratings of 5 for different list lengts, resulting from the SVD predictions.}
 %\label{fig:percentage_ratings_5}
%\end{figure}
%%%%%%%%%%%%%%%%

\smallskip
\noindent\textbf{The dataset sparsity affects accuracy.} We note that neighboring-based methods have lower accuracies. This is expected as these algorithms are notorious for suffering from sparsity. This effect is further amplified for UserNN as the average median number of interactions is three; hence, finding similar neighbors is more challenging. CoClustering also suffers from sparsity as the clustered entities are likely dissimilar because the similarity between users and items with few ratings cannot be determined reliably. The GNN, however, overcomes this challenge as it gathers information from multi-hop neighbors and learns the weighting parameter of each neighborhood from the data. Likewise, factorization-based approaches perform better as these suffer less from sparsity than neighboring-based ones when no strict cold starters are present.

%%%%%%%%%%%%%%%%
%\begin{figure}
%%\begin{figure}{R}{0.375\textwidth}
%\begin{center}
%  \includegraphics[width=.5\linewidth]{Figures/GNDCGvsNDCG.pdf}
%\end{center}
%\caption{Re-ranking tradeoff for the different algorithms. Each marker indicates a value of $\alpha$ in \eqref{eq.utility} from $\{0, 0.1, 0.2 \ldots, 1\}$.}
% \label{fig:nudging_weighted_results}
%\end{figure}

\begin{figure}[!tp]
\centering
\begin{subfigure}{.33\textwidth}
  \centering
  \includegraphics[width=1\linewidth]{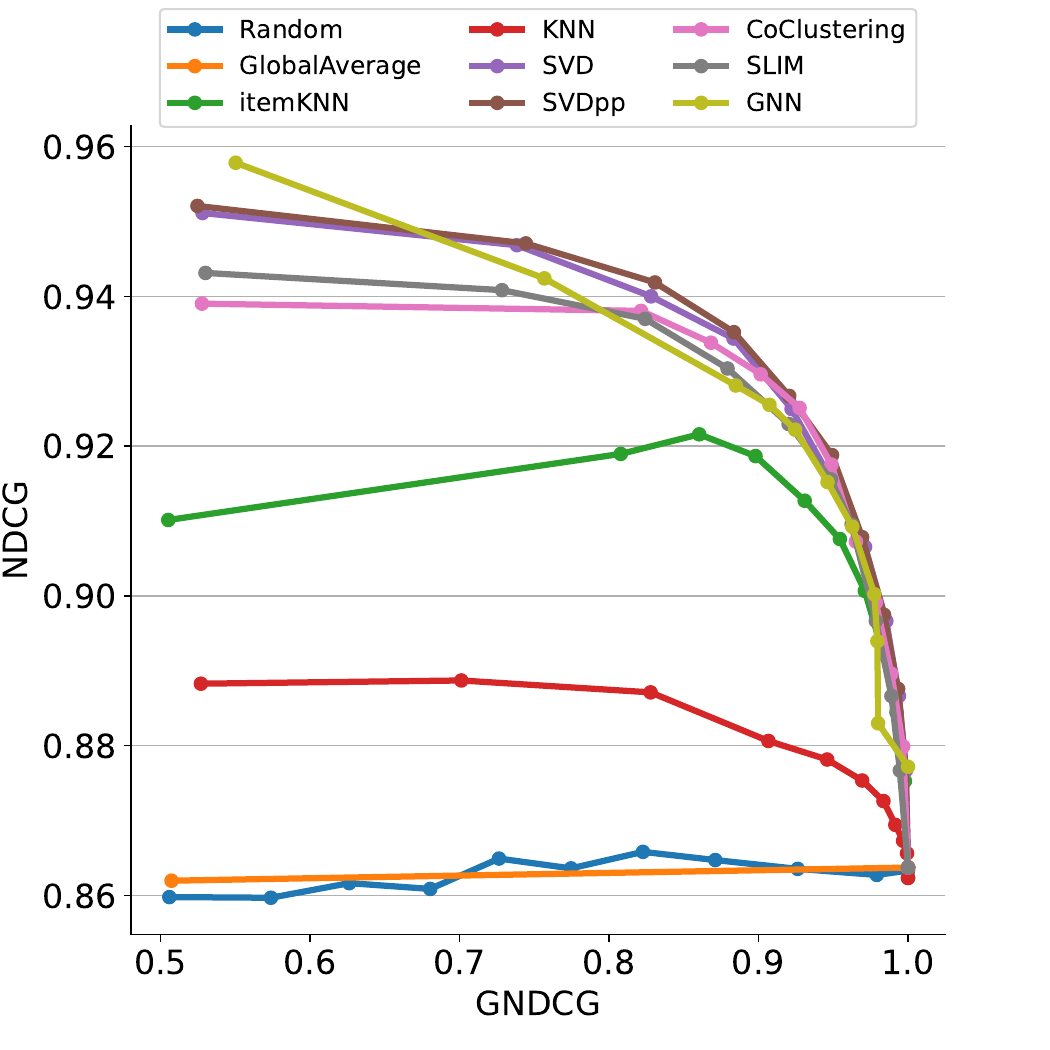}
  \caption{}
  \label{fig:tradeoff_CoClustering}
\end{subfigure}
\begin{subfigure}{.33\textwidth}
  \centering
  \includegraphics[width=1\linewidth]{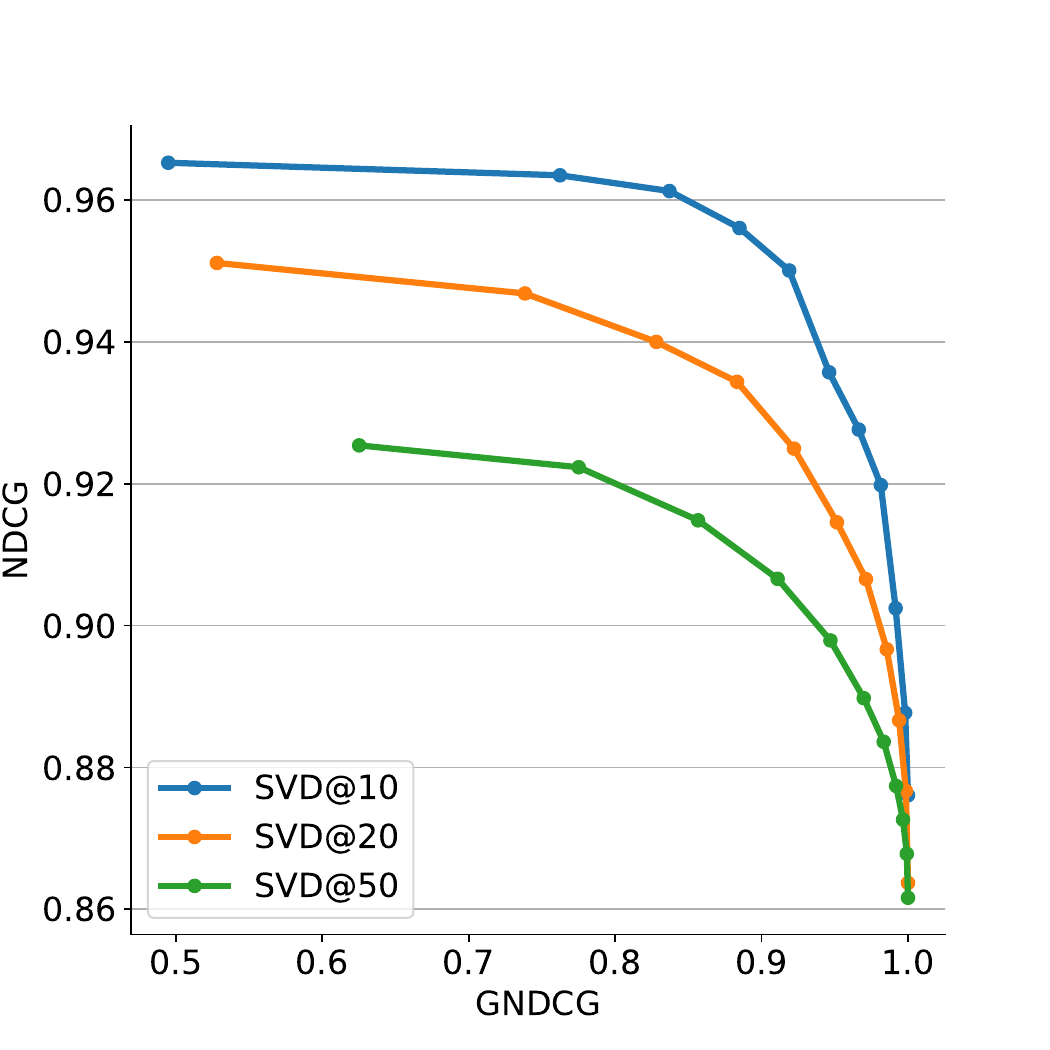}
  \caption{}
  \label{fig:tradeoff_SVD}
\end{subfigure}%
\begin{subfigure}{.33\textwidth}
  \centering
  \includegraphics[width=1\linewidth]{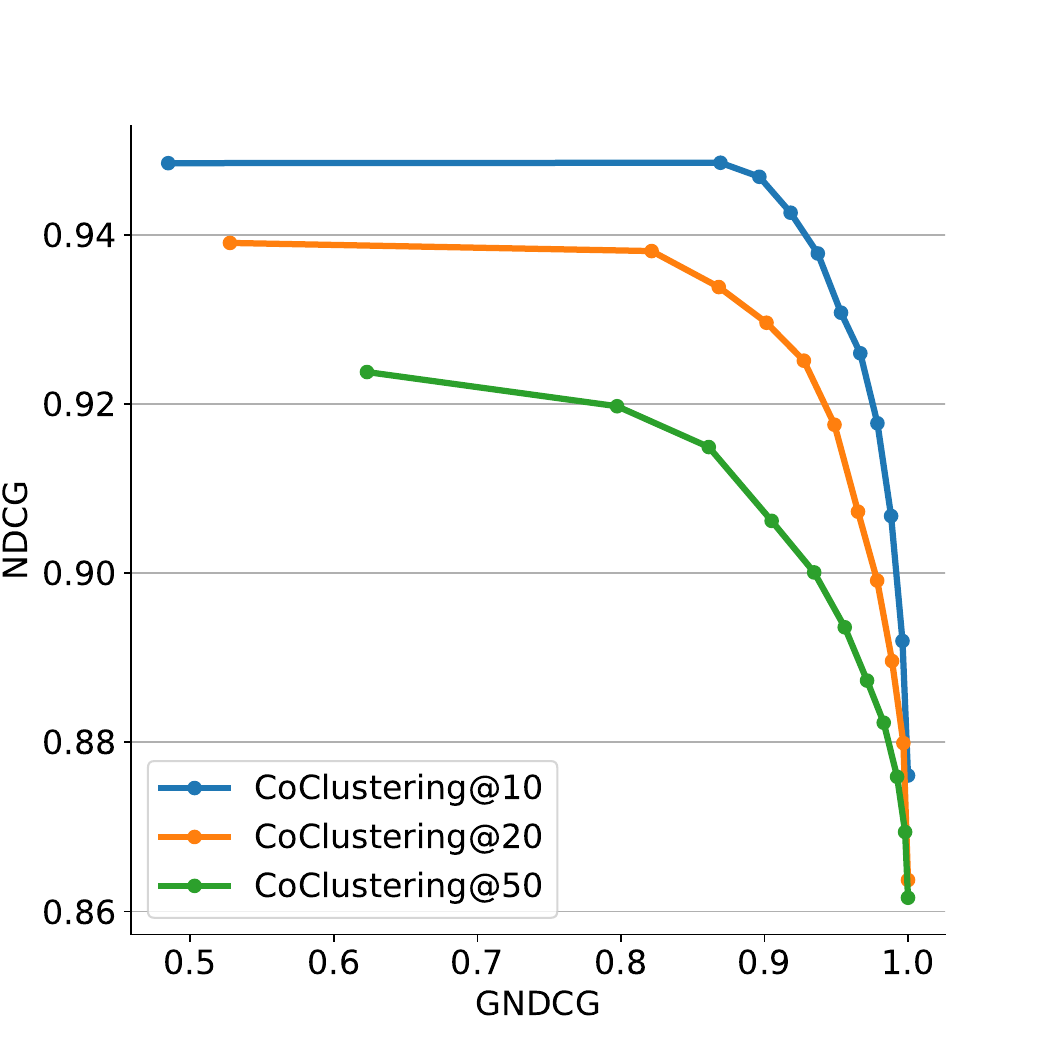}
  \caption{}
  \label{fig:tradeoff_CoClustering}
\end{subfigure}
\caption{Accuracy-greenness trade-off. Each marker indicates a value of $\alpha$ in \eqref{eq.utility}, chosen from $\{0, 0.1, 0.2 \ldots, 1\}$, where the leftmost marker represents the result for $\alpha=1$ and the rightmost marker represents the result for $\alpha=0$. \textbf{(a)} Reranking trade-off for the different algorithms. By changing $\alpha$, a small degree of accuracy is sacrificed for a larger gain in greenness. \textbf{(b-c)} Trade-off for different list lengths of the SVD and CoClustering algorithms. Longer lists lead to a worse performance for both metrics.}
%\caption{GNDCG-NDCG tradeoff for different list lengths of the SVD and CoClustering algorithms.}
\label{fig:tradeoff_list_lengths}
\end{figure}

%%%%%%%%%%%%%%%%

\subsection{Accuracy-Greenness Trade-off}\label{subsec:accGreen_trade}

Inspired by the literature on fairness-aware~\cite{suhr2019two,mansoury2021graph}, health~\cite{trattner2017investigating}, multi-objective~\cite{zheng2022survey,jannach2023survey}, and calibrated~\cite{steck2018calibrated} recommender systems, we now consider a reranking approach to improve the accuracy-greenness trade-off of recommender systems. We merge the estimated rating $\hat{r}_{u,i}$ with the item greenness $g_i$ into the utility score
\begin{equation}\label{eq.utility}
\mu_{u,i} = \alpha\hat{r}_{u,i} + (1-\alpha)g_i
\end{equation}
and rerank the items based on it. The scalar $\alpha \in [0,1]$ controls the trade-off: for $\alpha \to 1$ the utility is governed by the rating and the greenness plays a minor role in approaching the benchmarked performance; for $\alpha \to 0$ the utility is governed by the greenness score. Under the same experimental setting, we evaluate the accuracy-greenness trade-off as a function of $\alpha$ and list length $k$. Fig.~\ref{fig:tradeoff_list_lengths} (a) illustrates the results for a list length of $20$, whereas Fig.~\ref{fig:tradeoff_list_lengths} (b) and (c) show the impact of the list length for two representative algorithms. Again, We make three key observations.

\smallskip
\noindent\textbf{Reranking improves greenness, minimal accuracy loss.} By changing $\alpha$, we improve greenness at the expense of accuracy. However, the greenness growth rate is much higher than the accuracy drop rate. For instance, for the SVD, an $\alpha = 0.6$ improves the GNDCG@$20$ by $72.6\%$ whereas the NDCG$@20$ reduces by only $2.7\%$ w.r.t. the benchmarked results. The highest improvements in GNDCG and lowest decrease in NDCG are attained for values of $\alpha \ge 0.5$, suggesting that we can pay little in terms of accuracy to gain much in greenness and keep user satisfaction high. In fact, for $\alpha < 0.5$ the greenness outweighs accuracy in \eqref{eq.utility}; thus, there is a lower impact on the reranking as the different greenness values preserve the same relative order in the list. We should note that the accuracy-greenness relation and its impact on user satisfaction may not always be a trade-off. For instance, users who prioritize environmental concerns may prefer greener recommendations and rate them higher, thus achieving alignment between accuracy and greenness. On the other hand, if a greener recommendation is more expensive, consumers who prioritize saving money may be dissatisfied with the recommendation and give it a lower rating. We discuss these higher-level aspects in Sec.~\ref{subsec:accGreen_trade_Disc}, whereas the findings of Fig.~\ref{fig:tradeoff_list_lengths} show the potential of a simple, yet effective, method to improve list greenness of any RecSys algorithm.

\smallskip
\noindent\textbf{Factorization methods achieve a better trade-off.} This is because of their improved ability to generalize to all data and their ability to handle sparsity better than neighboring-based methods. This ensures a consistent higher accuracy while improving greenness. The neighboring-based methods show a different pattern. For the itemKNN algorithm, its accuracy increases between $\alpha = 1$ and $\alpha = 0.8$. We attribute this to their wrong rating estimates; 
%(see Table~\ref{tab_stat}); 
hence, when the greenness outweighs the estimated ratings, these miss-estimations become less important. 

\smallskip
\noindent\textbf{Longer lists affect the trade-off.} %Overall longer lists lead to a worse trade-off curve. 
By fixing one of the metrics (resp. NDCG or GNDCG) we see that the other one (resp. GNDCG or NDCG) reduces for longer list lengths. This is emphasized for $\alpha \ge 0.5$ that leads to a lower reduction of the GNDCG as a function of the list length $k$ (fixed the accuracy). This is because of the bigger weight given to the greenness score. However, for a fixed $\alpha$ we see that as the list length grows substantially (e.g., from $k = 20$ to $k = 50$) the NDCG reduces, whereas the GNDCG increases slightly, which is consistent with the observations made in Table.~\ref{tab_ratPerf} that corresponds to $\alpha = 1$ (leftmost marker).

\section{Discussion}\label{sec:discussion}

We proposed the first complete pipeline to research the potential of recommender system algorithms for recommending green choices. As a first step towards achieving such goal, we mined the first dataset that includes CO$_2$-eq as features for the items in addition to real user-item interactions, user features, and item features. Consequently, we benchmarked several popular RecSys algorithms in terms of accuracy and greenness. We found that no algorithm is greener than others, as they all ignore the greenness when making predictions. We then proposed a reranking strategy to promote greener recommendations by trading accuracy. This strategy improved the greenness of the benchmarked algorithms by up to +89.1\% while affecting accuracy in the worst case by {around 11\%}. We believe this work paths a new dimension within recommender system research and leads to several follow-up directions that we detail in the sequel.

\subsection{Data}\label{subsec:disc_dataset}

We built and made available the RecipeEmission dataset to overcome the main bottleneck of researching green recommender systems~\cite{rolnick2022tackling}. RecipeEmission contains the carbon footprint of items in kilograms of CO$_2$-eq as feature, along with a corresponding greenness score projected into a range of $[0,5]$, as defined by Eq.~\eqref{eq_greenness}. RecipeEmission presents long-tail distributions and statistics similar to conventional RecSys datasets (see Fig.~\ref{fig:dataset_props} and Table~\ref{tab_stat}). One distribution shows that most users interact with items with a low carbon footprint, and another shows that most items have a low carbon footprint. We also observed that user preferences were not based on the items' environmental impact.

The CO$_2$-eq should serve as a quality indicator for the item's carbon footprint rather than representing a precise value. This is because of the challenges in mining the carbon footprint of the different items, as well as of the uncertainties related to it (see Sec.~\ref{subsec:buildData}), both of which have been identified in earlier literature \cite{muller2012ingredient,angelsen2023healthiness}. Thus, we advocate using the greenness of $[0, 5]$ to measure the environmental impact of an item. If more precise footprints are needed, it is paramount, in the first place, to standardize the CO$_2$-eq of all items, accounting for their densities, dynamics, and seasonality of the CO$_2$-eq, as well as the uncertainties thereof. In a broader perspective, this requires quantifying the CO$_2$-eq of the entire process (e.g., material, production, shipping, incentives). This task is not only challenging from an engineering perspective, but it also needs the willingness of the retailers to collect and release such data .This may not be needed for building and assessing RecSys algorithms, and resorting to projected greenness values could lead to quicker development.

RecipeEmission is far from being an industrial-scale dataset. One way to improve scalability is to leverage the approach in \cite{belletti2019scaling} and generate a synthetic larger dataset. However, this will come with the additional bias of the CO$_2$-eq (or greenness) from RecipeEmission. In general, generating synthetic datasets by accounting for the greenness perspective poses interesting research challenges. It would also be possible to use natural language processing-based techniques to connect recipe and sustainability information \cite{van2021using}. Moreover, further research is needed to mine other real datasets and corroborate whether the distributions in Fig.~\ref{fig:co_rank_rating} are specific to this setting or also hold elsewhere. %\textcolor{magenta}{EI: this last sentence can be used as a point to address/answer one of the reviewer concerns that our findings may be dataset specific!!I.e., that we recognize such an issue and put it into discussion.}

\subsection{Benchmarking}\label{subsec:benchmark}

We analyzed different conventional RecSys algorithms in terms of accuracy and greenness. We observed that all algorithms ignore greenness and that longer lists are greener. In part, these findings are affected by the distributions of the greenness and ratings present in the dataset, as both contain a bias towards higher values. %\textcolor{magenta}{EI: this last sentence and the paragraph below can be used as a point to address/answer one of the reviewer concerns that our findings may be dataset specific!! I.e., that we recognize such an issue and put it into discussion.}

These observations should serve as a baseline and need extensive further investigation to be definitive for three main reasons. First, they have a dataset bias. It is commonly observed in RecSys that methods vary in performance in different settings. However, building green-based RecSys datasets is challenging, as highlighted in the previous section. One way to mitigate this is via offline A/B testing on the existing dataset \cite{gilotte2018offline}. Second, we used the predicted ratings to rank the items in the list. Our rationale was to avoid the dependency on any ranking-based criterion. Optimizing the methods with other criteria may provide additional insights on benchmarking; a popular choice could be the Bayesian Personalized Ranking \cite{rendle2012bpr}. Lastly, we evaluated the methods by ranking subsets of 100 user-item pairs each. This was done to contrast the dataset bias towards high rating values, but its downside is that ranked subsets could provide different results from those of a full ranked set. 

Future research could focus on developing additional metrics and assessing a broader range of algorithms. Particularly, we could develop metrics inspired by diversity or novelty \cite{castells2021novelty} to measure the greenness in the list. Regarding algorithms, it could be interesting to answer the question \emph{do deep learning-based RecSys algorithms recommend greener items than conventional ones}? We explicitly focused here on more conventional solutions to provide baselines for the more advanced ones but also because there is evidence that they compare well with the more advanced solutions \cite{ferrari2019we}. In this regard, we hypothesize that deep learning-based RecSys algorithms are \emph{less} green than conventional algorithms as they tend to overparameterize the model and overfit on accuracy. 

\subsection{Accuracy-Greenness Trade-off}\label{subsec:accGreen_trade_Disc}

nspired by the literature on fairness-aware recommender systems~\cite{mehrotra2018towards,mansoury2021graph} and calibrated recommendation~\cite{steck2018calibrated}, we proposed a reranking strategy to improve the accuracy-greenness trade-off for an algorithm in top-$k$ recommendations. This strategy is modular to any RecSys algorithm and ready to use. Our findings suggest that we can substantially improve the list greenness with little effect on accuracy and that longer lists lead to a worse trade-off.
There are three aspects worth discussing. First, the trade-off is controlled by a single scalar. Manually controlling this parameter could be challenging as its value depends on the RecSys algorithm, user, and application. Automatizing the trade-off as a multi-objective recommendation task~\cite{zheng2022survey,jannach2023survey} represents an important future direction. 
Second, the observed trade-off analysis is implicitly influenced by the dataset bias towards high ratings and low CO$_2$-eq values. Thus, it is easy for an algorithm to make recommendations greener in this setting, without compromising accuracy. The latter may not necessarily apply elsewhere. 
Third, we only studied the accuracy-greenness trade-off. Future studies could also take a more multidimensional trade-off (accuracy, diversity, novelty, serendipity), which could better align user satisfaction with list greenness. This relates to how we defined performance, which is limited to the explicit rating. By this reasoning, we considered items with an equal rating equally accurate, which is not true in practice.
An interesting future work is investigating the effectiveness of other reranking approaches on the accuracy-greenness trade-off of the recommendation models, such as the ones in~\cite{antikacioglu2017post,oneto2020general,singh2018fairness,zehlike2017fa}.

\subsection{Broader Impact}\label{subsec:broader_impact}

Fully understanding the potential of recommender systems remains challenging both because of the technical challenges and also because of the user involvement side. Adding to this, the sustainability aspects make it even more challenging. First, quantifying the greenness of an item is a rather complex task. It involves assessing where and how the item is produced, where the first material is obtained, and by what means the final item has been delivered to the user. All of these factors contribute to an item's CO$_2$-eq. For example, distance itself could play a major role as a similar item could be recommended by a closer provider. At the same time, sustainability in online shopping questions fundamental user principles. Users may be willing to pay for more expensive but eco-friendly items or choose slower delivery routes to contribute to the climate cause. Retailers, at the same time, may provide incentives and change marketing strategies to steer users towards greener choices.

One of our main conclusions is that recommending greener items comes at the expense of accuracy. However, this aspect needs to be questioned as the trade-off may not hold for user satisfaction. This is because our analysis did not consider the price of the items or purpose-driven users. In online shopping, price plays an important role. Users willing to pay more may be more satisfied with expensive but higher-quality eco-friendly items. At the same time, purpose-driven users could willingly opt for a greener recommendation to fulfill their internal drives. In both these cases, recommending a greener item may even lead to a higher accuracy and not necessarily to a trade-off. All these aspects call for a holistic approach when deploying RecSys in online shopping, which, on the one hand, poses engineering challenges to enable them to provide green recommendations and, on the other hand, poses business-consumers dilemmas. We believe that the findings of this work represent an important step towards this ambitious goal.

% %%%%%%
% %	Methods
% \section{Methods}\label{sec:methods}
% \input{methods.tex}

%%%%%%
%	Related work
\section{Related work}\label{sec:relatedwork}
% \section{E-commerce, Carbon Footprint and Recommender Systems}\label{appendix}
% \section{Related work}

%\textcolor{magenta}{EI: Some reviewers suggested works on energy and RecSys or other related works. If added here, we shall highlight them in blue.}

The carbon footprint of e-commerce has been analyzed by the academic community, and investigative journalism, as well as the major industrial actors that aim to promote their environmental-friendly strategies \cite{weideli2013environmental,van2014growth,rai2021net,panzone2021sustainable,cheah2022comparative,buldeo2022not}. User studies have also shown a willingness among consumers to prioritize environmental considerations, even at the expense of higher costs. In this context, we highlight some key findings that underscore the importance of studying carbon-aware RecSys.

\smallskip\noindent
\textbf{Carbon figures.} The carbon footprint of e-commerce is largely governed by four factors: i) transportation logistics; ii) packaging; iii) returns, and iv) direct operations \cite{ECommerce_Logistics}. Transportation, including last-mile delivery fleet, logistics, and freight contributes to $14.1$ MMT of CO$_2$-eq in the Chinese e-commerce market alone. E-commerce logistics contributed to 19 MMT CO$_2$-eq and this figure is forecasted to 25 MMT by 2030 \cite[pp 12]{WEFUrban}. 

Packaging, contributes to 8.71 MMT ($\sim23\%$) of CO$_2$-eq in China, and up to $45\%$ of total emissions worldwide. However, it has seen little change in the past decade despite the increased awareness from both sellers and consumers \cite{fantozzi2019life,escursell2021sustainability}. A 2019 study by Oceana reported that Amazon plastic packaging amounted to 211 million kg \cite[pp 12]{Oceania}, and this is expected to increase further due its improved sales \cite{AmazonSales}. 

Return rates account for a surprising 25\% of the total emissions. In the United States alone, the emissions resulting from the transportation of returns increased from 15 MMT CO$_2$-eq in 2019 to 27 MMT CO$_2$-eq in 2021 \cite[pp 5]{Optoro}. Return rates for online purchased products vary per country and age, ranging from 6-8.5\% for senior users to 10-21\% for young adults \cite{AlvarezShape}. Even considering the lowest return rate of 6\%, this amounts to more than 3.47 billion return packages operated by U.S. Amazon and in the Chinese e-commerce market\footnote{We used the information from 2020 as reported by Amazon~\cite{AmazonSales} and the Chinese State Post Bureau as reported by South China Morning Post (an Alibaba group owned newspaper) \cite{SouthChina}.}. As item return rates are linked to user satisfaction with the purchased product, and the latter is affected by the RecSys algorithm of the retailer, we believe that recommendations could play a significant role in improving the carbon footprint of this voice. The Amazon RecSys algorithm, for example, contributes to 35\% of consumer purchases \cite{McKinsey} and could be optimized to reduce the number of returns.

Lastly, online browsing is another factor contributing to the e-commerce carbon footprint. For instance, data centers' operation in the Chinese market accounts for $11$ MMT of CO$_2$-eq emissions. In the United Kingdom, the top-eight online fast-fashion retailers generated approximately 40 grams of CO$_2$-eq per website visit in 2021 \cite{UKFashion}. Based on data from similarweb.com, we estimate that these eight websites generated around 2.6 metric tons of CO$_2$-eq only from online visits of United Kingdom-based consumers. Extrapolating these figures to the Amazon setting leads to around 135,000 metric tons of CO$_2$-eq only by online visits\footnote{We considered an average of $5$ grams of CO$_2$-eq per visit and an average of 2.5 billion worldwide visits per month estimated from \cite{AmazonVisits}.}. We believe that RecSys algorithms could be leveraged to personalize recommendations and limit the number of returned visits, thus reducing the carbon footprint of e-commerce.

\smallskip\noindent
\textbf{User reaction.} Recent research and surveys have highlighted that sustainability is becoming a driver for consumer purchasing decisions. The recent reports in \cite{Shopify2022,SendCloud2023} revealed that one in two consumers perceives online shopping as harmful to the environment, and 44\% them are willing to purchase from brands that have a clear commitment to sustainability. Additionally, one in four online shoppers is willing to pay extra for sustainable products and shipping \cite{PostNord2021}. Another survey by Mercado Libre showed that 12\% of their users prioritize the durability and recycled materials of the products \cite{MercadoLibre2022}. Sustainable packaging is another critical factor that influences online brand choices, as evidenced by several surveys \cite{PackagingUSUK, PackagingUS, PackagingUK, xie2021assessing}. In the United States, packaging concerns were a driving factor for one in four consumers, whereas in the United Kingdom, one in three consumers were influenced by sustainable packaging. The Mercado Libre survey also found that sustainable packaging concerns drove the purchasing decisions of 28\% of their users. The recent work in \cite{wang2023downside} studied the user reaction when a tablet was either recommended by and AI algorithm or a human. It concluded that users showed a lower preference for green products when it was recommended by an AI algorithm.

\smallskip\noindent
\textbf{RecSys studies.} The above figures demonstrate the potential of RecSys for a greener e-commerce. In fact, RecSys principles have been implemented in sustainability-focused research \cite{ross2010collaborative,ibrahim2023recommendation}, which utilize collaborative filtering to calculate individual carbon emissions based not only on personal metadata but also community correlations. The role of RecSys algorithms in promoting energy-efficient behavior in buildings was surveyed in \cite{himeur2021survey,knijnenburg2014smart,starke2017effective}. The study suggests that incorporating explanations, visualization, and time-aware information in recommendations improves their quality and acceptance among users, as well as their satisfaction. RecSys algorithms have also been applied to sustainable travel planning \cite{bothos2015recommender}. In this work, a system ranks the potential routes based on a utility value that takes into account factors such as total duration, walking or cycling time, comfort, and CO$_2$ emissions. The study involved 28 participants, who were positive about their recommendations. The work in \cite{anagnostopoulos2021predictive} utilized RecSys algorithms to predict potential commuting partners for users engaging in shared mobility. The study showed promising results in predicting preferences and improving the user satisfaction. The work in \cite{lee2011recommender} proposed a RecSys architecture idea for adaptive electronic green marketing based on fuzzy logic but its usefulness has not been tested nor compared with baselines. Finally, the work in \cite{tomkins2018sustainability} proposed a probabilistic framework that utilizes domain knowledge to identify sustainable products and users, and predict future purchases. However, none of the aforementioned works conducted a thorough comparison on the role of RecSys in promoting green recommendations, nor did they study the tradeoff between recommendation accuracy and greenness. This work targeted both of these gaps.

%%%%%%
%	Conclusion
\section{Conclusion}\label{sec:conclusion}
This paper addresses the critical gap in evaluating the environmental impact of recommender systems by introducing a dataset with carbon footprint emissions and exploring how RecSys algorithms can promote more sustainable choices. Our research highlights the significant carbon footprint associated with online shopping and demonstrates that traditional RecSys algorithms, while optimized for accuracy, often overlook the greenness of recommendations. We discovered that longer recommendation lists tend to be greener but at the cost of accuracy. We proposed a straightforward, yet powerful, reranking approach that integrates carbon footprint data to enhance the sustainability of recommendations while maintaining a reasonable level of accuracy. This modular reranking method can be seamlessly applied to existing RecSys algorithms without requiring model retraining, offering a practical solution for improving recommendation sustainability. Looking ahead, future work will involve investigating the effectiveness of various reranking solutions from the literature in addressing the accuracy-greenness trade-off. Additionally, we aim to enhance the confidence and quality of our data collection process and explore the applicability of our methods to other recommendation domains.

% \begin{appendices}
% \input{Appendix_converstion.tex}
% \end{appendices}

\appendix
%%%%
\section{Details of recommendation algorithms}\label{appendix}

In this section, we describe the recommendation algorithms and hyperparameter values used for running experiments in this paper.

\begin{itemize}
    \item \textbf{Random predictions} This method generates random predictions for all items in the test set.
    \item \textbf{Global mean.} This method computes the global mean rating of the training set and assigns that value to the items in the test set.
    \item \textbf{Item nearest neighbor (ItemNN).} This is a neighboring-based method using item-item collaborative filtering. The cosine similarity is used to find the neighboring items. We estimate the predicted rating $\hat{r}_{ui}$, between an item $i$ and a user $u$ as:
\begin{equation}
\hat{r}_{u i}=\frac{\sum_{j \in \ccalN^{k}(i)} \operatorname{sim}(i, j)  r_{u j}}{\sum_{v \in \ccalN^{k}(i)} \operatorname{sim}(i, j)}
    \label{eq:knn}
\end{equation}
where $\ccalN^{k}(i)$ is the set of $k$ neighbors to item $i$. We considered the number of neighbors in \{20, 40, 60\}.
    \item \textbf{User nearest neighbor (UserNN).} This is the complementary collaborative filter of the ItemNN. We followed the same setup.
    \item \textbf{SVD.} This low-rank approach approximates the user-item rating matrix and uses these latent representations to fill the missing ratings. For a user $u$ and item $i$, we use the corresponding latent feature vectors, $\mathbf{p}_u$ and $\mathbf{q}_i$ to approximate a rating $\hat{r}_{ui}$ as:
\begin{equation}
    \hat{r}_{ui} = b_u + b_i + \mathbf{p}_u^\top\mathbf{q}_i.
    \label{eq:IRSVD}
\end{equation}
We optimized for $\{20, 50, 100, 150\}$ factors, a learning rate of $\{0.1, 0.01, 0.001, 0.0001\}$, a regularization parameter of $\{0.1, 0.01, 0.001, 0.0001\}$ and training epochs of $\{20, 50, 80\}$. The optimization was done using stochastic gradient descent.
    \item \textbf{SVD++.} This enhances the SVD by accounting for implicit ratings, representing that users rated items in the first place, regardless of the rating value. Let the set of items rated by user $u$ be $\ccalN(u)$ and denote the latent features also as $\bbq_i$ and $\bbp_u$. Let also $\bby_j$ be the implicit feedback for item $j \in \ccalN(u)$. Then, we predict the ratings as:
\begin{equation}
\hat{r}_{u i}=\mu+ b_u+b_i+\mathbf{q}_{i}^{T}\left(\mathbf{p}_{u}+|\ccalN(u)|^{-\frac{1}{2}} \sum_{j \in \ccalN(u)} \bby_{j}\right)
\label{eq:SVDpp}
\end{equation}
where $|\cdot|$ is the set cardinality operator. We optimized for the same parameters as regular SVD. 
    \item \textbf{CoClustering.} This algorithm assigns all users and items to clusters (containing either users or items) and co-clusters (containing both users and items). Then, it uses the means of these clusters and co-clusters to predict the ratings. We used $k-$means clustering and predicted the ratings as:
\begin{equation}
\hat{r}_{u i}=\overline{C_{u i}}+\left(\mu_{u}-\overline{C_{u}}\right)+\left(\mu_{i}-\overline{C_{i}}\right)
\label{eq:CoClustering}
\end{equation}
with $\overline{C_{u i}}$ being the average rating of co-cluster $C_{u i}$, and $\overline{C_{u}}$ and $\overline{C_{i}}$ being the average ratings of clusters $C_u$ and $C_i$ respectively. We optimized for $\{3, 6, 12, 24\}$ user and item (co-)clusters and for $\{20, 50, 80\}$ epochs.
    \item \textbf{Slim.} This is a linear optimization-based model using the ElasticNet principle to find sparse neighbors for each user and item. The estimated rating is computed as follows:

\begin{equation}
\hat{r}_{u i}=\mathbf{r}_{u}^{\top} \mathbf{w}_{i}
\label{eq:slim1}
\end{equation}
where $\mathbf{w}_{i}$ is a sparse aggregation coefficient vector. Using matrices, this model can be expressed as:

\begin{equation}
\mathbf{\hat{R}}=\mathbf{R} \mathbf{W}
\label{eq:slim2}
\end{equation}
where $\bbR$ is the user-item rating matrix and $\bbW$ collects all aggregation coefficients. To estimate the matrix $\mathbf{W}$, the following problem is solved:

\begin{equation}\label{eq:slim3}
\begin{split}
\underset{{\mathbf{W}}}{\operatorname{minimize}} \;\; & \frac{1}{2}\|\mathbf{R}-\mathbf{R W}\|_{F}^{2}+\frac{\beta}{2}\|\mathbf{W}\|_{F}^{2}+\lambda\|\mathbf{W}\|_{1} \\
\text { subject to } \;\; & \mathbf{W} \geq 0 \\
& \operatorname{diag}(\mathbf{W})=0
\end{split}
\end{equation}

% \begin{equation}
% \begin{array}{ll}
% \underset{{\mathbf{W}}}{\operatorname{minimize}} & \frac{1}{2}\|\mathbf{R}-\mathbf{R W}\|_{F}^{2}+\frac{\beta}{2}\|\mathbf{W}\|_{F}^{2}+\lambda\|\mathbf{W}\|_{1} \\
% \text { subject to } & \mathbf{W} \geq 0 \\
% & \operatorname{diag}(\mathbf{W})=0
% \end{array}
% \label{eq:slim3}
% \end{equation}

\noindent where the matrix $\ell_1$ norm is computed elementwise. We optimized for the regularization parameters $\beta$ and $\lambda$ in $\{0.005, 0.05, 0.5\}$.
    \item \textbf{Graph neural networks.} We considered one of the first approaches used for RecSys matrix completion that is based on graph convolutions \cite{berg2017graph}. The user-item rating matrix $\bbR$ is transformed into an undirected bipartite graph connecting users to items. The edge weights of this graph correspond to the provided rating value. Then, the matrix completion task is transformed into a link prediction problem, which is solved via a graph convolution autoencoder. This autoencoder leverages convolutional GNNs to learn node-specific embeddings and subsequently decodes the latter to predict the links. We optimized for a learning rate of {0.01, 0.05, 1}, dropout ratios of {0.3, 0.5, 0.7}, and {20, 50, 80} epochs.
\end{itemize}

\bibliographystyle{ACM-Reference-Format}
\bibliography{sample-base}

% \appendix
% \input{Appendix_converstion}

\end{document}